\documentclass[aps,prd,reprint,twocolumn,10pt,eqsecnum,showpacs,nofootinbib,amsfonts,amssymb,amsmath,longbibliography,superscriptaddress,floatfix]{revtex4-1}
\usepackage[colorlinks=true,allcolors=blue]{hyperref}
\usepackage{bm}
\usepackage{xcolor}
\usepackage{mathtools}

\DeclareSymbolFont{bbold}{U}{bbold}{m}{n}
\DeclareSymbolFontAlphabet{\mathbbold}{bbold}

\newcommand{\ud}{\mathrm{d}}
\newcommand{\uD}{\mathrm{D}}
\newcommand{\pb}[1]{\,\mbox{}_{#1}}
\newcommand{\pp}[1]{\,\mbox{}^{#1}}
\newcommand{\lie}{\pounds}
\newcommand{\vkappa}[1]{\overset{_\varkappa}{#1}}
\newcommand{\zero}[1]{\overset{_{\scalebox{0.4}{0}}}{#1}}
\newcommand{\half}[1]{\overset{_{\scalebox{0.4}{1/2}}}{#1}}

\newcommand{\smallunderset}[2]{\underset{^{#1}}{#2}}

\newcommand{\mat}[1]{\underline{#1}}

\usepackage[normalem]{ulem}

\colorlet{RED}{red}
\colorlet{OLIVE}{olive}

\begin{document}

\author{\'Eanna \'E.\ Flanagan}
\email{eef3@cornell.edu}
\affiliation{Department of Physics, Cornell University, Ithaca, New York, 14853, USA}
\author{Alexander M.\ Grant}
\email{amg425@cornell.edu}
\affiliation{Department of Physics, Cornell University, Ithaca, New York, 14853, USA}
\author{Abraham I.\ Harte}
\email{abraham.harte@dcu.ie}
\affiliation{Centre for Astrophysics and Relativity, School of Mathematical Sciences Dublin City University, Glasnevin, Dublin 9, Ireland}
\author{David A.\ Nichols}
\email{david.nichols@virginia.edu}
\affiliation{Department of Physics, University of Virginia, P.O.~Box 400714,
Charlottesville, Virginia 22904-4714, USA}
\affiliation{Gravitation Astroparticle Physics Amsterdam (GRAPPA), University of Amsterdam, Science Park, P.O.~Box 94485, 1090 GL Amsterdam, The Netherlands}

\title{Persistent gravitational wave observables: Nonlinear plane wave spacetimes}

%
%
\newcount\hh
\newcount\mm
\mm=\time
\hh=\time
\divide\hh by 60
\divide\mm by 60
\multiply\mm by 60
\mm=-\mm
\advance\mm by \time
\def\hhmm{\number\hh:\ifnum\mm<10{}0\fi\number\mm}

\begin{abstract}
  In the first paper in this series, we introduced ``persistent gravitational wave observables'' as a framework for generalizing the gravitational wave memory effect.
  These observables are nonlocal in time and nonzero in the presence of gravitational radiation.
  We defined three specific examples of persistent observables: a generalization of geodesic deviation that allowed for arbitrary acceleration, a holonomy observable involving a closed curve, and an observable that can be measured using a spinning test particle.
  For linearized plane waves, we showed that our observables could be determined just from one, two, and three time integrals of the Riemann tensor along a central worldline, when the observers follow geodesics.
  In this paper, we compute these three persistent observables in nonlinear plane wave spacetimes, and we find that the fully nonlinear observables contain effects that differ qualitatively from the effects present in the observables at linear order.
  Many parts of these observables can be determined from two functions, the \emph{transverse Jacobi propagators}, and their derivatives (for geodesic observers).
  These functions, at linear order in the spacetime curvature, reduce to the one, two, and three time integrals of the Riemann tensor mentioned above.
\end{abstract}

\maketitle



\tableofcontents

\section{Introduction}

In a previous paper~\cite{Flanagan2019}, we introduced a class of ``persistent gravitational wave observables'' that generalize gravitational wave memory effects.
We included three specific examples of persistent observables: a generalization of geodesic deviation to allow for accelerated curves, which we called \emph{curve deviation}; an observable involving the solution of a particular differential equation along a closed curve, which was a type of \emph{holonomy}; and a collection of observables involving the separation of a spinning test particle from an observer, as well as the particle's momentum and intrinsic spin.
We then explicitly computed these three observables for perturbations off of flat spacetime.
Assuming that these perturbations represented gravitational plane waves, we found that our observables (assuming unaccelerated observers) could be written in terms of just one, two, and three time integrals of the Riemann tensor along a central worldline.

Gravitational wave memory effects are a special class of persistent observables.
In~\cite{Flanagan2019}, we defined memory effects to be persistent observables that are associated with boundary symmetries and conservation laws at spacetime boundaries.
The boundary can be future null infinity (as in the case of the initial investigations of the memory effect~\cite{Zeldovich1974, Christodoulou1991}; see also the recent review~\cite{Strominger2017} and references therein for later developments) or the event horizon of a black hole~\cite{Hawking2016, Compere2018b, Chandrasekaran2018}.
This paper will not focus on spacetime boundaries and will instead consider persistent observables in spacetime interiors.

In this paper, we will consider the persistent observables defined in~\cite{Flanagan2019} in the context of exact, nonlinear plane wave spacetimes.
While we explored our observables in~\cite{Flanagan2019} to linear order in the curvature of the spacetime near the observers, the nonlinear properties of these observables are less well understood.
Exact plane wave spacetimes provide a simple context in which to study these nonlinear properties, as demonstrated in other work discussing persistent observables in the literature (see, for example,~\cite{Zhang2017a, Zhang2017b, Zhang2018, Harte2015}).

The effects we will compute in this paper will be nonlinear in the amplitude of the gravitational waves.
Gravitational waves produced by astrophysical sources, however, will be weak when the waves have reached any detector,  so effects that are nonlinear in the amplitude of the gravitational wave are not expected to be detectable by current detectors.\footnote{Note that the nonlinear memory effect~\cite{Christodoulou1991} is \emph{not} nonlinear in the amplitude of the gravitational wave at the detector; rather, it arises from a nonlinearity in Einstein's equations in asymptotically flat spacetimes.
It is much more likely to be detected by current and future gravitational wave detectors~\cite{Lasky2016, Islo2019, Hubner2019, Boersma2020}.}
Nevertheless, these effects are qualitatively different from linear effects, and therefore interesting in their own right.
There may also be regimes in which they are detectable by future detectors.

\subsection{Simplified model of geodesic deviation}

To illustrate the types of distinctive effects that can arise in persistent observables beyond the linearized approximation, we now discuss a simplified model of geodesic deviation.
Consider the following differential equation for a function $\xi(u)$:

\begin{equation} \label{eqn:toy_jacobi}
  \ddot{\xi} (u) = \epsilon \ddot{f} (u) \xi(u),
\end{equation}
where $\epsilon \ll 1$, and dots denote derivatives with respect to $u$.
This is a scalar version of the geodesic deviation equation, where $\xi$ is the separation between two observers and $\epsilon f$ is the equivalent of the gravitational wave strain amplitude.
Consider the analogue of a burst of gravitational waves that occurs between $u = 0$ and $u = U$, where $\ddot{f} (u) = 0$ for all $u < 0$ and $u > U$.
For simplicity, set $f(u) = 0$ for $u < 0$ [in general, this will imply that $f(u) \neq 0$ for $u > U$].
The solution for $\xi(u)$ at some time $u > 0$ is then given by

\begin{equation} \label{eqn:toy_soln}
  \xi(u) = a(u) \xi(0) + b(u) \dot{\xi} (0),
\end{equation}
where

\begin{equation} \label{eqn:toy_ics}
  a(0) = \dot{b} (0) = 1, \qquad \dot{a} (0) = b(0) = 0.
\end{equation}
Therefore, to second order in $\epsilon$,

\begin{equation} \label{eqn:toy_soln_a}
  \begin{split}
    a(u) = 1 &+ \epsilon f(u) + \frac{1}{2} \epsilon^2 [f(u)]^2 \\
    &- \epsilon^2 \int_0^u \ud u' \int_0^{u'} \ud u'' [\dot{f} (u'')]^2 + O(\epsilon^3).
  \end{split}
\end{equation}
The counterpart of the first-order memory in this case is given by the term in~\eqref{eqn:toy_soln_a} linear in $\epsilon$ [that is, $f(u)$ after the burst].
This function is at most linear in $u$, since $\ddot{f} (u)$ is zero at late times.
However, even if this first-order memory is zero, that is, if $f(u) = 0$ for $u > U$, the second-order memory is nonzero and would, in general, grow linearly with time:

\begin{equation}
    a(u > U) = Cu + D,
\end{equation}
where

\begin{subequations}
  \begin{align}
    C &\equiv -\epsilon^2 \int_0^\infty \ud u' [\dot{f} (u')]^2 + O(\epsilon^3) \neq 0, \\
    D &\equiv \epsilon^2 \int_0^\infty \ud u' \int_{u'}^\infty \ud u'' [\dot{f} (u'')]^2 + O(\epsilon^3) \neq 0.
  \end{align}
\end{subequations}
Since the coefficient $C$ is nonzero, observers in this simplified model would have a relative velocity after the burst: at second order all nontrivial solutions must have $\dot{a} (u) \neq 0$ after the burst.
At first order, there is no such restriction on the final relative velocity, so first- and second-order calculations yield qualitatively different results.
While Eq.~\eqref{eqn:toy_jacobi} is only a simplified model of geodesic deviation, the explicit discussion given in Sec.~\ref{sec:2nd_jacobi} is qualitatively similar.
For example, nonlinear plane wave spacetimes always have a nonzero relative velocity after a burst, often called ``velocity memory''~\cite{Bondi1989, Harte2015, Zhang2017a}.

Another motivation for considering nonlinear plane wave spacetimes is as follows.
Our persistent observables are ``degenerate'' in the linearized, plane wave limit, in the sense that they can be written in terms of only three functions (one, two, and three time integrals of the Riemann tensor), in the case where the observers are unaccelerated~\cite{Flanagan2019}, even though the form of the observables allows them to have more nonzero, independent components than these three functions possess.
This implies that while our observables can encode a wide range of qualitatively different physical effects, the effects are all determined by the same, limited set of properties of the gravitational wave.
One might expect that at higher order these degeneracies are broken.
However, we instead find in this paper that these degeneracies (or linear relationships between observables) are replaced with nonlinear relationships between observables.

An example of such a nonlinear relationship occurs in the simplified model~\eqref{eqn:toy_jacobi}: it can be shown from Eq.~\eqref{eqn:toy_jacobi} that the Wronskian

\begin{equation} \label{eqn:toy_wronskian}
  W = a(u) \dot{b} (u) - \dot{a} (u) b(u)
\end{equation}
must be conserved, and by Eq.~\eqref{eqn:toy_ics}, we have $W = 1$.
This holds to all orders in $\epsilon$; however, one can use Eq.~\eqref{eqn:toy_ics} to show that

\begin{equation}
  \begin{gathered}
    \dot{a} (u) = O(\epsilon), \quad a(u) = 1 + \int_0^u \ud u' \dot{a} (u'), \\
    b(u) = u + O(\epsilon),
  \end{gathered}
\end{equation}
from which Eq.~\eqref{eqn:toy_wronskian} becomes

\begin{equation} \label{eqn:toy_degeneracy}
  \dot{b} (u) - 1 + \int_0^u \ud u' [\dot{a} (u') - \dot{a} (u)] = O(\epsilon^2).
\end{equation}
The quantity~\eqref{eqn:toy_degeneracy} is an example of a combination of observables that vanishes at first order in the curvature (corresponding to a degeneracy), but is nonzero at higher orders.
This example, moreover, shows that some relationships that hold at first order are approximations to fully nonlinear relationships between observables.
Much of this paper focuses upon finding and understanding these nonlinear relationships.

\subsection{Results}

In this paper, we consider the three persistent gravitational wave observables of~\cite{Flanagan2019} in plane wave spacetimes: the curve deviation, holonomy, and spinning test particle observables.
The values of these three observables in plane wave spacetimes are given in Eq.~\eqref{eqn:curve_dev_results} for curve deviation, Eq.~\eqref{eqn:spinning_results} for the spinning test particle observables, and Appendix~\ref{app:holonomy} for the holonomy observable.
These expressions rely upon a fair amount of notation defined throughout; for an overview, see Table~\ref{tab:symbols} at the end of this paper.

A key result of this paper is that the curve deviation and holonomy observables, when considered for geodesic curves, can be determined \emph{exactly} in plane wave spacetimes, to all orders in initial separation and relative velocity.
In particular, they can be written in terms of two sets of functions and their first derivatives, which are analogous to the functions $a(u)$ and $b(u)$ of Eq.~\eqref{eqn:toy_soln}, as well as their first derivatives.
These functions are the transverse components of the Jacobi propagators (defined in Sec.~\ref{sec:bitensors} below), which have been extensively studied in these spacetimes~\cite{Harte2012a, Harte2012b, Harte2015}.
The information needed to construct these transverse Jacobi propagators can be obtained by measuring the displacement memory (leading and subleading~\cite{Flanagan2019}) and its time derivative, the relative velocity observable, in these spacetimes.
It is known that other quantities in plane wave spacetimes, such as solutions to the geodesic equation, can be written in terms of these transverse Jacobi propagators as well~\cite{Harte2012a}.
The transverse Jacobi propagators and their derivatives form a set of three (and not four) independent matrix functions, because of a constraint analogous to Eq.~\eqref{eqn:toy_wronskian}.

Some of the observables of~\cite{Flanagan2019} we compute perturbatively instead of exactly.
The first of these are our curve deviation and holonomy observables for nongeodesic curves.
We find that these observables can be expressed as time integrals involving the transverse Jacobi propagators, but \emph{cannot} be expressed locally in time in terms of these propagators and their time derivatives, as they can be for geodesic curves.
Roughly, this is because these observables can be written as integrals involving the product of the transverse Jacobi propagators and a given, but arbitrary acceleration vector.
The other observable we calculate perturbatively is the persistent observable arising from a spinning test particle.
Here again, it does not seem possible to express this observable locally in time in terms of products and derivatives of transverse Jacobi propagators, likely because this observable is \emph{also} defined in terms of an accelerating curve.
Observables which cannot be written locally in time in terms of products and derivatives of transverse Jacobi propagators measure features of the gravitational waves that are independent of the leading and subleading displacement memory and the relative velocity observables.

\subsection{Summary and conventions}

The structure of the paper is as follows.
First, in Sec.~\ref{sec:review}, we review the properties of plane wave spacetimes and the forms that the ``fundamental bitensors'' (parallel and Jacobi propagators) take in these spacetimes (results that were obtained in~\cite{Harte2012a}).
We also introduce the transverse Jacobi propagators mentioned above, and we provide some intuition for these functions by computing them in detail at second order in the curvature and for a particular plane wave spacetime.
In Sec.~\ref{sec:observables}, we review the persistent observables that we introduced in~\cite{Flanagan2019} and give explicit formulas for these observables in plane wave spacetimes.
We present further discussion and our conclusions in Sec.~\ref{sec:discussion}.

Throughout this paper, we use the same conventions as those in~\cite{Flanagan2019}: the conventions for the metric and curvature tensors given in Wald~\cite{Wald1984} and the conventions for bitensors from Poisson's review article~\cite{Poisson2004} (we use a slightly different convention for coincidence limits from what is used in~\cite{Poisson2004}; see Footnote~\ref{fn:coinc_limit}).
We use lowercase Latin letters from the beginning of the alphabet ($a$, $b$, etc.) for abstract spacetime tensor indices; for abstract tensor indices on the linear and angular momentum bundle (see~\cite{Flanagan2016, Flanagan2019}), we use the corresponding uppercase letters ($A$, $B$, etc.).
For convenience, we are using a convention for bitensors where we use the same annotations for indices as are used on the points at which the indices apply (e.g., $a$, $b$, etc. at the point $x$ and $a'$, $b'$, etc. at the point $x'$).
We will omit the arguments of bitensors when no ambiguity arises due to the annotation of indices.
For example, $g^{a'}{}_a (x', x)$ will be abbreviated as $g^{a'}{}_a$, whereas $\sigma^a (x', x)$ will only be abbreviated as $\sigma^a (x')$, and $\sigma(x', x)$ will not be abbreviated at all.
Finally, for brevity, we will occasionally  take powers of order symbols, writing,
for example, $O(a, b)^3$ as shorthand for $O(a^3, a^2 b, a b^2, b^3)$.

\section{Review of Plane Wave Spacetimes} \label{sec:review}

In this section, we review properties of exact, nonlinear plane wave spacetimes.
These are spacetimes with metrics that can be written, in Brinkmann coordinates $(u, v, x^1, x^2)$~\cite{Brinkmann1925}, as

\begin{equation} \label{eqn:metric}
  \ud s^2 = -2 \ud u \ud v + \mat{\mathcal A}{}_{ij} (u) x^i x^j \ud u^2 + \ud x^i \ud x^j \delta_{ij},
\end{equation}
where $u$ is the phase of the gravitational wave, and $\underline{\mathcal A}{}_{ij} (u)$ is the wave profile.\footnote{Another coordinate system, Rosen coordinates~\cite{Rosen1937}, is often used in these spacetimes.
  This coordinate system is the nonlinear generalization of TT gauge for linearized gravity; see, for example, ~\cite{Harte2015} for more details.}
The particular signs and constant factors that have been chosen in this metric are the same as those in~\cite{Harte2012a}.
Our convention for tensor components in Brinkmann coordinates is that we use $u$ and $v$ as indices for $u$ and $v$ components, and we use lowercase Latin letters from the middle of the alphabet ($i$, $j$, etc.) for the remaining two components, which we will call the \emph{transverse} components.
When considering generic components in Brinkmann coordinates, we use lowercase Greek letters from the middle of the alphabet ($\mu$, $\nu$, etc.).
For these component indices, we use the Einstein summation convention.
Tensors which are only nonzero in their transverse ($i$, $j$, etc.) components we denote with underlines, and refer to as being transverse.

We now list several basic features of these spacetimes which we will need in this paper (for a review, see~\cite{Ehlers1962}).
The first is the existence of a null vector field $\ell^a$ which is covariantly constant:

\begin{equation} \label{eqn:constant}
  \nabla_a \ell^b = 0.
\end{equation}
In terms of Brinkmann coordinates, this vector field is given by

\begin{equation} \label{eqn:l}
  \ell^a \equiv -(\partial_v)^a
\end{equation}
(note that our convention for $\ell^a$ is that of~\cite{Ehlers1962}, which differs from that of~\cite{Harte2012a} by a sign).
We also define an antisymmetric tensor

\begin{equation} \label{eqn:volume}
  \underline{\epsilon}{}_{ab} \equiv 2 (\ud x^1)_{[a} (\ud x^2)_{b]}.
\end{equation}
This tensor is transverse, and is a volume form on surfaces of constant $u$ and $v$.
Finally, the Riemann tensor in plane wave spacetimes is given by

\begin{equation}
  R_{abcd} = 4 \ell_{[a} \underline{\mathcal A}{}_{b][c} \ell_{d]},
\end{equation}
where

\begin{equation} \label{eqn:A_def}
  \underline{\mathcal A}_{ab} \equiv \underline{\mathcal A}_{ij} (u) (\ud x^i)_a (\ud x^j)_b.
\end{equation}
It then follows from $\underline{\mathcal A}{}_{ab} \ell^b = 0$ that the only constraint from Einstein's equations is that

\begin{equation}
  T_{ab} = -8\pi \underline{\mathcal A}^c{}_c \ell_a \ell_b.
\end{equation}
Therefore, in vacuum, $\underline{\mathcal A}^a{}_a = 0$.

\subsection{Geodesics and symmetries} \label{sec:geodesics}

We now discuss the solution of the geodesic and Killing equations in plane wave spacetimes.
Consider a geodesic $\gamma$, affinely parametrized by $\tau$ and with tangent vector $\dot{\gamma}^a$.
At a given value of $\tau$, we denote the coordinates of $\gamma(\tau)$ by $u$, $v$, and $x^i (\tau)$, and at $\tau'$, we denote the coordinates by $u'$, $v'$, and $x^i (\tau')$.\footnote{Note that $x^i (\tau')$ lacks a prime on the index $i$; this notation will be justified in Sec.~\ref{sec:bitensors}.}

We define the parameter

\begin{equation} \label{eqn:chi}
  \chi \equiv \dot{\gamma}^a \ell_a,
\end{equation}
which is conserved along the geodesic $\gamma$ by Eq.~\eqref{eqn:constant}.
This implies that

\begin{equation}
  u' = u + \chi (\tau' - \tau). \label{eqn:u_geo}
\end{equation}
Geodesics can be classified by whether $\chi$ vanishes.
For the case $\chi = 0$, the geodesic lies entirely within a surface of constant $u$, and one can show that

\begin{equation}
  \ddot{x}^i (\tau) = 0, \qquad \ddot{v} = 0;
\end{equation}
therefore, the solutions of the geodesic equation are linear functions of $\tau$.
For the case $\chi \neq 0$, the geodesic equation for $x^i (\tau)$ is given by

\begin{equation} \label{eqn:x_geo_diff}
  \ddot{x}^i (\tau) = \chi^2 \mat{\mathcal A}^i{}_j (u) x^j (\tau),
\end{equation}
which has nontrivial solutions.

The solutions to Eq.~\eqref{eqn:x_geo_diff} can be written in terms of two functions of $u$ and $u'$, $\mat{K}^i{}_j (u', u)$ and $\mat{H}^i{}_j (u', u)$, that satisfy the differential equations

\begin{subequations} \label{eqn:2d_jacobi}
  \begin{align}
    \partial_{u'}^2 \mat{K}^i{}_j (u', u) &= \mat{\mathcal A}^i{}_k (u') \mat{K}^k{}_j (u', u), \\
    \partial_{u'}^2 [(u' - u) \mat{H}^i{}_j (u', u)] &= (u' - u) \mat{\mathcal A}^i{}_k (u') \mat{H}^k{}_j (u', u),
  \end{align}
\end{subequations}
with the boundary conditions

\begin{subequations} \label{eqn:2d_BCs}
  \begin{align}
    \mat{K}^i{}_j (u, u) &= \mat{H}^i{}_j (u, u) = \delta^i{}_j, \\
    \left.\partial_{u'} \mat{K}^i{}_j (u', u)\right|_{u' = u} &= \left.\partial_{u'} \mat{H}^i{}_j (u', u)\right|_{u' = u} = 0
  \end{align}
\end{subequations}
(see, for example,~\cite{Harte2012a}).
We call these functions the \emph{transverse Jacobi propagators}, since they are related to the transverse components of certain bitensors called Jacobi propagators (as we will discuss in Sec.~\ref{sec:bitensors}).
When we say that something in plane wave spacetimes is known ``exactly,'' we mean that it can be written in terms of $\mat{K}^i{}_j (u', u)$ and $\mat{H}^i{}_j (u', u)$.
The solution to Eq.~\eqref{eqn:x_geo_diff} is then
\begin{equation} \label{eqn:x_geo}
  x^i (\tau') = \mat{K}^i{}_j (u', u) x^j (\tau) + (\tau' - \tau) \mat{H}^i{}_j (u', u) \dot{x}^j (\tau),
\end{equation}
where on the right-hand side $u'$ and $u$ are determined from $\tau$ and $\tau'$ by Eq.~\eqref{eqn:u_geo}.

Next, to solve for $v'$ when $\chi \neq 0$, for convenience we assume that $\gamma$ is timelike.
Note that

\begin{equation}
  \psi^a = -2v \ell^a + x^i (\partial_i)^a
\end{equation}
is a proper homothety, satisfying $\lie_\psi g_{ab} = 2 g_{ab}$ (see, for example,~\cite{Maartens1991}).
As a consequence of this~\cite{Patino1993}, it follows that $\dot{\gamma}^a \psi_a + \tau$ is conserved along $\gamma$, so one can write $v'$ in terms of the coordinate $v$ of $\gamma(\tau)$:

\begin{equation} \label{eqn:v_geo}
  v' = v - \frac{1}{2\chi} \left[x^i (\tau') \dot{x}_i (\tau') - x^i (\tau) \dot{x}_i (\tau) + (\tau' - \tau)\right].
\end{equation}
In the above equation, one could use the values of $x^i (\tau')$ and $\dot{x}^i (\tau')$ that were determined in Eq.~\eqref{eqn:x_geo} in order to write everything in terms of $\tau$, $\tau'$, transverse Jacobi propagators, and initial data.
The normalization $\dot{\gamma}^a \dot{\gamma}_a = -1$ implies that

\begin{equation} \label{eqn:four_velocity}
  \begin{split}
    \dot{\gamma}^a = &\chi (\partial_u)^a + \dot{x}^i (\tau) (\partial_i)^a \\
    &- \frac{1}{2\chi} \left[1 + \dot{x}^i (\tau) \dot{x}_i (\tau) + \mat{\mathcal A}{}_{ij} (u) x^i (\tau) x^j (\tau)\right] \ell^a,
  \end{split}
\end{equation}
which is consistent with Eq.~\eqref{eqn:v_geo}.

The quantities $\mat{K}^i{}_j (u', u)$ and $\mat{H}^i{}_j (u', u)$ are also useful for finding Killing vectors in plane wave spacetimes~\cite{Harte2012a}.
Plane wave spacetimes possess a four-parameter family of Killing vector fields in addition to $\ell^a$~\cite{Ehlers1962}.
We denote a member of this family by

\begin{equation} \label{eqn:killing}
  \Xi^a \equiv -x^i \dot{\mat \Xi}{}_i (u) \ell^a + \mat{\Xi}^i (u) (\partial_i)^a,
\end{equation}
where the function $\mat{\Xi}^i (u)$ is any solution to

\begin{equation}
  \mat{\ddot \Xi}{}^i (u) = \mat{\mathcal A}^i{}_j (u) \mat{\Xi}^j (u).
\end{equation}
The value of this function at any initial phase $u_0$ determines its values at any other $u$:

\begin{equation} \label{eqn:killing_ics}
  \mat{\Xi}^i (u) = \mat{K}^i{}_j (u, u_0) \mat{\Xi}^j (u_0) + (u - u_0) \mat{H}^i{}_j (u, u_0) \mat{\dot \Xi}{}^j (u_0).
\end{equation}
Since $\mat{\Xi}^i (u_0)$ and $\mat{\dot \Xi}{}^i (u_0)$ are four numbers, the space of Killing vectors of the form~\eqref{eqn:killing} is four-dimensional.

Finally, we list a few useful properties of the transverse Jacobi propagators $\mat{K}^i{}_j (u', u)$ and $\mat{H}^i{}_j (u', u)$: first, Eq.~\eqref{eqn:2d_jacobi} implies (see, for example,~\cite{Harte2012a})~\footnote{For arbitrary solutions $\mat{K}^i{}_j (u', u)$ and $\mat{H}^i{}_j (u', u)$ to Eqs.~\eqref{eqn:2d_jacobi} (that is, ignoring boundary conditions), we note that the quantity in Eq.~\eqref{eqn:wronskian} is independent of $u$ and $u'$.
  One can think of this quantity as a conserved symplectic form on the space of solutions to Eqs.~\eqref{eqn:2d_jacobi}~\cite{Grasso2018}, and Eqs.~\eqref{eqn:2d_jacobi} form a Hamiltonian system~\cite{Uzun2018}.}

\begin{equation} \label{eqn:wronskian}
  \begin{split}
    \mat{K}{}_k{}^i &(u', u) \partial_{u'} \left[(u' - u) \mat{H}^k{}_j (u', u)\right] \\
    &- (u' - u) \mat{H}^k{}_j (u', u) \partial_{u'} \mat{K}{}_k{}^i (u', u) = \delta^i{}_j.
  \end{split}
\end{equation}
This relationship is an analogue of Eq.~\eqref{eqn:toy_wronskian} and shows that there are only \emph{three} independent quantities among $\underline{K}^i{}_j (u', u)$, $\underline{H}^i{}_j (u', u)$, $\partial_{u'} \underline{K}^i{}_j (u', u)$, and $\partial_{u'} \underline{H}^i{}_j (u', u)$.
One can also show the following relationships hold when these two propagators' arguments are switched~\cite{Harte2012b}:

\begin{subequations} \label{eqn:switch_identities}
  \begin{align}
    \mat{H}^i{}_j (u', u) &= \mat{H}{}_j{}^i (u, u'), \\
    \partial_{u'} \mat{K}^i{}_j (u', u) &= -\partial_u \mat{K}_j{}^i (u, u').
  \end{align}
\end{subequations}
Finally, using the fact that derivatives of the transverse Jacobi propagators with respect to their second argument also must satisfy Eq.~\eqref{eqn:2d_jacobi}, one has that~\cite{Harte2012b}

\begin{subequations} \label{eqn:deriv_identities}
  \begin{gather}
    \partial_u \mat{K}^i{}_j (u', u) = -(u' - u) \mat{H}^i{}_k (u', u) \mat{\mathcal A}^k{}_j (u), \\
    \partial_u \left[(u' - u) \mat{H}^i{}_j (u', u)\right] = -\mat{K}^i{}_j (u', u).
  \end{gather}
\end{subequations}
These identities are quite useful for deriving the results in Sec.~\ref{sec:observables}.

\subsection{Parallel and Jacobi propagators} \label{sec:bitensors}

In this section, we provide explicit expressions for the parallel and Jacobi propagators, which are the bitensors that are needed for calculating the persistent observables of~\cite{Flanagan2019}.
These bitensors are most naturally expressed in terms of the transverse Jacobi propagators defined in Sec.~\ref{sec:geodesics} above.

We first review the definitions of the parallel and Jacobi propagators in arbitrary spacetimes.
The parallel and Jacobi propagators are one-forms at $x \equiv \gamma(\tau)$ and vectors at $x' \equiv \gamma(\tau')$, and are defined to be solutions of the following differential equations along $\gamma$: the parallel propagator $\pb{\gamma} g^{a'}{}_a$ obeys

\begin{equation} \label{eqn:parallel}
  \frac{\uD}{\ud \tau'} \pb{\gamma} g^{a'}{}_a = 0
\end{equation}
(where $\uD/\ud \tau' \equiv \dot{\gamma}^{a'} \nabla_{a'}$), whereas the Jacobi propagators $\pb{\gamma} K^{a'}{}_a$ and $\pb{\gamma} H^{a'}{}_a$ obey

\begin{subequations} \label{eqn:jacobi}
  \begin{align}
    \frac{\uD^2}{\ud \tau'^2} &\pb{\gamma} K^{a'}{}_a \nonumber \\
    &= -R^{a'}{}_{c'b'd'} \dot{\gamma}^{c'} \dot{\gamma}^{d'} \pb{\gamma} K^{b'}{}_a, \\
    \frac{\uD^2}{\ud \tau'^2} &\left[(\tau' - \tau) \pb{\gamma} H^{a'}{}_a\right] \nonumber \\
    &= -(\tau' - \tau) R^{a'}{}_{c'b'd'} \dot{\gamma}^{c'} \dot{\gamma}^{d'} \pb{\gamma} H^{b'}{}_a.
  \end{align}
\end{subequations}
Note that we are using the notation described in the Introduction, where indices at $x$ are denoted by $a$, $b$, etc., whereas at $x'$ they are denoted by $a'$, $b'$, etc.
The following boundary conditions are imposed for these differential equations~\footnote{\label{fn:coinc_limit} These boundary conditions are given in the language of coincidence limits (see~\cite{Poisson2004}; note that our notation for coincidence limits is more explicit than this reference, as we use subscripts to indicate the limit being taken).
  We use the standard convention (for example, in~\cite{Poisson2004}) where indices at the point whose limit is being taken ($x'$ in this case) are treated as if they were at the limiting point ($x$) for expressions that occur outside of the coincidence limit.}:

\begin{subequations}
  \begin{align}
    \left[\pb{\gamma} g^{a'}{}_b\right]_{\tau' \to \tau} &= \left[\pb{\gamma} K^{a'}{}_b\right]_{\tau' \to \tau} = \left[\pb{\gamma} H^{a'}{}_b\right]_{\tau' \to \tau} \nonumber \\
    &= \delta^a{}_b, \\
    \left[\frac{\uD}{\uD \tau'} \pb{\gamma} K^{a'}{}_b\right]_{\tau' \to \tau} &= \left[\frac{\uD}{\uD \tau'} \pb{\gamma} H^{a'}{}_b\right]_{\tau' \to \tau} = 0.
  \end{align}
\end{subequations}
These bitensors $\pb{\gamma} g^{a'}{}_a$, $\pb{\gamma} K^{a'}{}_a$, and $\pb{\gamma} H^{a'}{}_a$ are defined for a given curve $\gamma$ which connects the points $x$ and $x'$.
In the case where two points $x$ and $x'$ lie within a convex normal neighborhood (that is, are close enough that there is a unique geodesic connecting them), the parallel and Jacobi propagators that are defined in terms of this unique geodesic are denoted simply by $g^{a'}{}_a$, $K^{a'}{}_a$, and $H^{a'}{}_a$.
For most of this paper (except for in Appendix~\ref{app:acceleration}) we will restrict attention to geodesic curves $\gamma$.
However, even in this case, it will sometimes be necessary to specify this curve $\gamma$ in the propagators $\pb{\gamma} g^{a'}{}_a$, $\pb{\gamma} K^{a'}{}_a$, and $\pb{\gamma} H^{a'}{}_a$.
This is because $\gamma$ may not be the only geodesic between $x$ and $x'$, when $\tau' - \tau$ is sufficiently large, due to the occurrence of caustics (see~\cite{Harte2012a} for a discussion of caustics in these spacetimes).
To allow for the existence of caustics, we specify the curve $\gamma$ explicitly when it is needed.

In plane wave spacetimes, the parallel and Jacobi propagators can be given in exact form when the curve $\gamma$ along which they are computed is a geodesic, as shown in~\cite{Harte2012a}.
This can be done by finding a convenient basis $(e_\alpha)^a$ at $\gamma(\tau)$ and constructing a basis $(e_\alpha)^{a'}$ at $\gamma(\tau')$, either using parallel transport,

\begin{equation} \label{eqn:parallel_basis}
  \frac{\uD}{\ud \tau} (e_\alpha)^a = 0
\end{equation}
(for the parallel propagator), or using the Jacobi equation,

\begin{equation} \label{eqn:jacobi_basis}
  \frac{\uD^2}{\ud \tau^2} (e_\alpha)^a = -R^a{}_{cbd} \dot{\gamma}^c \dot{\gamma}^d (e_\alpha)^b
\end{equation}
(for the Jacobi propagators).
The parallel and Jacobi propagators can then be constructed from such a basis and its corresponding dual basis.
This method is similar to that of~\cite{Marck1983}, which was used to determine the parallel propagator in the Kerr spacetime.
Two of the basis elements are given by $\dot{\gamma}^a$ and $\ell^a$, which automatically satisfy Eqs.~\eqref{eqn:parallel_basis} and~\eqref{eqn:jacobi_basis}.
For brevity, we do not give the full details of this calculation.

Before we give the results of this calculation, we note that one result is that

\begin{equation} \label{eqn:2d_parallel}
  \pb{\gamma} g^{i'}{}_i = (\partial_j)^{i'} (\ud x^j)_i,
\end{equation}
by inspection of the connection coefficients of the metric~\eqref{eqn:metric}.
That is, the ``transverse parallel propagator'' is trivial.
To simplify expressions in this paper, we will no longer annotate the transverse indices $i$, $j$, etc. with primes in our expressions in Brinkmann coordinates, since distinguishing between primed and unprimed components is not necessary in view of Eq.~\eqref{eqn:2d_parallel}.
However, since these indices no longer indicate the point at which the bitensor is being evaluated, we will explicitly indicate the dependence on this point, which for many of the bitensors will be a dependence on proper time or $u$.
For example, instead of writing $\pb{\gamma} K^{i'}{}_i$, we will write $\pb{\gamma} K^i{}_j (\tau', \tau)$, and $\pb{\gamma} K^{i'}{}_u$ will be written as $\pb{\gamma} K^i{}_u (\tau')$.
This notation is consistent with the fact that we referred to the $x^i$ coordinates of $\gamma(\tau)$ and $\gamma(\tau')$ by $x^i (\tau)$ and $x^i (\tau')$, respectively, in Sec.~\ref{sec:geodesics}.

The values of the parallel and Jacobi propagators are different based on whether the parameter $\chi$ is zero or nonzero.
When $\chi = 0$ and spacetime is flat, one can show that

\begin{equation} \label{eqn:constant_u}
  \pb{\gamma} g^{a'}{}_a = \pb{\gamma} K^{a'}{}_a = \pb{\gamma} H^{a'}{}_a = (\partial_\mu)^{a'} (\ud x^\mu)_a,
\end{equation}
where, as mentioned above, $x^\mu$ refers to the $\mu$th Brinkmann coordinate.
When $\chi \neq 0$ and $\gamma$ is timelike, the nonzero components of the parallel and Jacobi propagators are~\cite{Harte2012a}

\begin{subequations} \label{eqn:4d_propagators}
  \begin{align}
    \pb{\gamma} g^{u'}{}_u &= \pb{\gamma} K^{u'}{}_u = \pb{\gamma} H^{u'}{}_u = 1, \\
    \pb{\gamma} g^{v'}{}_v &= \pb{\gamma} K^{v'}{}_v = \pb{\gamma} H^{v'}{}_v = 1, \\
    \pb{\gamma} g^i{}_j (\tau', \tau) &= \delta^i{}_j, \\
    \pb{\gamma} g^i{}_u (\tau') &= \frac{1}{\chi} \left[\dot{x}^i (\tau') - \dot{x}^i (\tau)\right], \\
    \pb{\gamma} g^{v'}{}_i (\tau) &= \frac{1}{\chi} \left[\dot{x}_i (\tau') - \dot{x}_i (\tau)\right], \\
    \pb{\gamma} g^{v'}{}_u &= \frac{1}{2\chi^2} \Big\{\left[\dot{x}^i (\tau') - \dot{x}^i (\tau)\right] \left[\dot{x}_i (\tau') - \dot{x}_i (\tau)\right] \nonumber \\
    &\hspace{3.5em}+ \chi \big[\underline{\mathcal A}{}_{ij} (u') x^i (\tau') x^j (\tau') \nonumber \\
    &\hspace{5.5em}- \underline{\mathcal A}{}_{ij} (u) x^i (\tau) x^j (\tau)\big]\Big\}, \displaybreak[0] \\
    \pb{\gamma} K^i{}_j (\tau', \tau) &= \mat{K}^i{}_j (u', u), \\
    \pb{\gamma} K^i{}_u (\tau') &= \frac{1}{\chi} \left[\dot{x}^i (\tau') - \mat{K}^i{}_j (u', u) \dot{x}^j (\tau)\right], \\
    \pb{\gamma} K^{v'}{}_i (\tau) &= \frac{1}{\chi} \left[\dot{x}_j (\tau') \mat{K}^j{}_i (u', u) - \dot{x}_i (\tau)\right], \\
    \pb{\gamma} K^{v'}{}_u &= \frac{1}{2\chi^2} \Big\{\dot{x}^i (\tau') \dot{x}_i (\tau') + \dot{x}^i (\tau) \dot{x}_i (\tau) \nonumber \\
    &\hspace{3.5em}- 2 \dot{x}_i (\tau') \mat{K}^i{}_j (u', u) \dot{x}^j (\tau) \nonumber \\
    &\hspace{3.5em}+ \chi \big[\mat{\mathcal A}{}_{ij} (u') x^i (\tau') x^j (\tau') \nonumber \\
    &\hspace{5.5em}- \mat{\mathcal A}{}_{ij} (u) x^i (\tau) x^j (\tau)\big]\Big\}, \displaybreak[0] \\
    \pb{\gamma} H^i{}_j (\tau', \tau) &= \mat{H}^i{}_j (u', u), \\
    \pb{\gamma} H^i{}_u (\tau') &= \frac{1}{\chi} \left[\dot{x}^i (\tau') - \mat{H}^i{}_j (u', u) \dot{x}^j (\tau)\right], \\
    \pb{\gamma} H^{v'}{}_i (\tau) &= \frac{1}{\chi} \left[\dot{x}_j (\tau') \mat{H}^j{}_i (u', u) - \dot{x}_i (\tau)\right], \\
    \pb{\gamma} H^{v'}{}_u &= \frac{1}{2\chi^2} \Big\{\dot{x}^i (\tau') \dot{x}_i (\tau') + \dot{x}^i (\tau) \dot{x}_i (\tau) \nonumber \\
    &\hspace{3.5em}- 2 \dot{x}_i (\tau') \mat{H}^i{}_j (u', u) \dot{x}^j (\tau) \nonumber \\
    &\hspace{3.5em}+ \chi \big[\mat{\mathcal A}{}_{ij} (u') x^i (\tau') x^j (\tau') \nonumber \\
    &\hspace{5.5em}- \mat{\mathcal A}{}_{ij} (u) x^i (\tau) x^j (\tau)\big]\Big\}.
  \end{align}
\end{subequations}
As in Eq.~\eqref{eqn:x_geo}, $u'$ and $u$ on the right-hand sides of these equations are functions of $\tau'$ and $\tau$ by Eq.~\eqref{eqn:u_geo}.
Note also that we have written the expressions in Eqs.~\eqref{eqn:4d_propagators} in terms of $x^i (\tau')$ and $\dot{x}^i (\tau')$, which can be expressed in terms of $x^i (\tau)$ and $\dot{x}^i (\tau)$ using Eq.~\eqref{eqn:x_geo}.

\subsection{Second-order transverse Jacobi propagators} \label{sec:2nd_jacobi}

We now compute general expressions for the transverse Jacobi propagators to second order in the curvature.
These results have been previously computed in~\cite{Harte2015}.
In the context of an arbitrary plane wave spacetime, one can write down perturbative expansions of the transverse Jacobi propagators in powers of $\mat{\mathcal A}^i{}_j (u)$:

\begin{subequations} \label{eqn:2d_jacobi_expansion}
  \begin{align}
    \mat{K}^i{}_j (u', u) &= \sum_{n = 0}^\infty \pp{(n)} \mat{K}^i{}_j (u', u), \\
    \mat{H}^i{}_j (u', u) &= \sum_{n = 0}^\infty \pp{(n)} \mat{H}^i{}_j (u', u).
  \end{align}
\end{subequations}
At zeroth order, from the boundary conditions in Eq.~\eqref{eqn:2d_BCs},
the transverse Jacobi propagators are

\begin{equation}
  \pp{(0)} \mat{K}^i{}_j (u', u) = \pp{(0)} \mat{H}^i{}_j (u', u) = \delta^i{}_j.
\end{equation}
Higher-order terms in this expansion are then obtained by solving Eqs.~\eqref{eqn:2d_jacobi} and~\eqref{eqn:2d_BCs} iteratively.
At first order, the propagators are given by

\begin{subequations}
  \begin{align}
    \pp{(1)} \mat{K}^i{}_j &(u', u) = \int_u^{u'} \ud u'' \int_u^{u''} \ud u''' \mat{\mathcal A}^i{}_j (u'''), \label{eqn:1st_jacobi_a} \\
    \pp{(1)} \mat{H}^i{}_j &(u', u) = \int_u^{u'} \ud u'' \int_u^{u''} \ud u''' \frac{u''' - u}{u' - u} \mat{\mathcal A}^i{}_j (u'''). \label{eqn:1st_jacobi_b}
  \end{align}
\end{subequations}
We write all higher-order corrections in terms of these first-order terms and their derivatives, as they provide a particularly convenient way of representing these results.
Note, however, that there is a certain amount of freedom in how we write second-order terms, because of the truncation of the identity~\eqref{eqn:wronskian} at first order.
As such, there are different ways of writing the first- and second-order results in this section, depending upon whether one uses all four of $\pp{(1)} \mat{K}^i{}_j (u', u)$, $\pp{(1)} \mat{H}^i{}_j (u', u)$, $\partial_{u'} \pp{(1)} \mat{K}^i{}_j (u', u)$, and $\partial_{u'} \pp{(1)} \mat{H}^i{}_j (u', u)$, or some subset of three.
As it results in relatively compact equations, we use all four.

Continuing to second order, one can show (by an integration by parts) that

\begin{widetext}
\begin{subequations} \label{eqn:2nd_jacobi}
  \begin{align}
    \pp{(2)} \mat{K}^i{}_j (u', u) &= \frac{1}{2} \pp{(1)} \mat{K}^i{}_k (u', u) \pp{(1)} \mat{K}^k{}_j (u', u) \nonumber \\
    &\hspace{1em}- \int_u^{u'} \ud u'' \int_u^{u''} \ud u''' \left\{\partial_{u'''} \pp{(1)} \mat{K}^i{}_k (u''', u) \partial_{u'''} \pp{(1)} \mat{K}^k{}_j (u''', u) - \frac{1}{2} \big[\mat{\mathcal A} (u'''), \pp{(1)} \mat{K} (u''', u)\big]{}^i{}_j\right\}, \label{eqn:2nd_jacobi_a} \\
    (u' - u) \pp{(2)} \mat{H}^i{}_j (u', u) &= \frac{1}{2} (u' - u) \pp{(1)} \mat{H}^i{}_k (u', u) \pp{(1)} \mat{H}^k{}_j (u', u)\nonumber \\
    &\hspace{1em}- \int_u^{u'} \ud u'' \int_u^{u''} \ud u''' \bigg\{(u''' - u) \partial_{u'''} \pp{(1)} \mat{H}^i{}_k (u''', u) \partial_{u'''} \pp{(1)} \mat{H}^k{}_j (u''', u) \nonumber \\
    &\hspace{10.5em}- \frac{1}{2} (u''' - u) \big[\mat{\mathcal A} (u'''), \pp{(1)} \mat{H} (u''', u)\big]{}^i{}_j\bigg\}, \label{eqn:2nd_jacobi_b}
  \end{align}
\end{subequations}
\end{widetext}
where the commutator $[A, B]^a{}_b$ is given by

\begin{equation}
  [A, B]^a{}_b \equiv A^a{}_c B^c{}_b - B^a{}_c A^c{}_b.
\end{equation}
Note that there are two types of terms that appear in Eqs.~\eqref{eqn:2nd_jacobi} at second order.
The first are terms that are merely squares of the final values of the first-order terms; these are the first terms in Eqs.~\eqref{eqn:2nd_jacobi_a} and~\eqref{eqn:2nd_jacobi_b}.
The other two terms in both equations are qualitatively different at second order.
They are generically nonzero, even when the final values of the first-order terms vanish, as they depend on integrals of the first-order terms throughout the curved region.
These terms are analogous to the fourth term in the simple model~\eqref{eqn:toy_soln_a} of the Introduction.

The different terms at second order are also qualitatively different in the following sense.
Assuming a vacuum plane wave, one has that $\mat{\mathcal A}^i{}_j (u)$ is traceless, and so $\pp{(1)} \mat{K}^i{}_j (u', u)$ and $\pp{(1)} \mat{H}^i{}_j (u', u)$ are as well.
Thus, we find that the first two terms in~\eqref{eqn:2nd_jacobi_a} and~\eqref{eqn:2nd_jacobi_b} are pure trace, as they are squares of $2 \times 2$ symmetric, trace-free matrices, and that the third terms are antisymmetric.
Because of the existence of pure trace terms at second order, gravitational waves possess an effective ``breathing'' polarization mode~\cite{Will2014} at this order~\cite{Harte2015}.
Note that the third (antisymmetric) term in Eq.~\eqref{eqn:2nd_jacobi_a} vanishes when the gravitational waves are linearly polarized; this effect was previously noted in~\cite{Zhang2018}.

\subsection{Example of a plane wave spacetime} \label{sec:model_waveform}

We now illustrate the general results of Sec.~\ref{sec:2nd_jacobi} by specializing to an explicit example of a plane wave spacetime.
We choose $\mat{\mathcal A}_{ij} (u)$ to vanish outside of the interval $[0, 2\pi n/\omega]$, where $n$ is a positive integer, and inside the interval we choose

\begin{equation} \label{eqn:model_waveform}
  \begin{split}
    \mat{\mathcal A}_{ij} (u) = \epsilon \omega^2 \bigg[&\sqrt{1 - a^2} \sin(\omega u) \pb{+} \mat{e}{}_{ij} \\
    &+ a \sin(\omega u + \phi) \pb{\times} \mat{e}{}_{ij}\bigg],
  \end{split}
\end{equation}
where

\begin{equation} \label{eqn:plus_cross}
  \pb{+} \mat{\boldsymbol e} = \begin{pmatrix}
    1 & 0 \\
    0 & -1
  \end{pmatrix}, \qquad \pb{\times} \mat{\boldsymbol e} = \begin{pmatrix}
    0 & 1 \\
    1 & 0
  \end{pmatrix}.
\end{equation}
This represents a wave pulse that contains $n$ full periods, and is a mixture of $+$ and $\times$ polarizations.\footnote{This wave profile is periodic, so in the fully nonlinear regime Floquet theory (see, for example,~\cite{Magnus1979} and references therein) applies to Eq.~\eqref{eqn:2d_jacobi} and its solutions.
  Although it is outside the scope of this paper, it would be interesting to use this fact to determine regions in the parameter space of $\epsilon$, $a$, and $\phi$ where solutions are bounded and regions in this parameter space where they are unbounded.}
Some special cases are linear polarization, where $\phi = 0$, and circular polarization, where $\phi = \pm \pi/2$ and $a = 1/\sqrt{2}$.
This wave pulse also satisfies $\int_{-\infty}^\infty \ud u \mat{\mathcal A}_{ij} (u) = 0$, which (at first order) means vanishing relative velocity at late times for observers that are initially comoving.
Gravitational waves at null infinity are also frequently assumed to satisfy a condition analogous to $\int_{-\infty}^\infty \ud u \mat{\mathcal A}_{ij} (u) = 0$.

Using the explicit wave profile in Eq.~\eqref{eqn:model_waveform}, we find that

\begin{widetext}
\begin{subequations} \label{eqn:2d_jacobi_model}
  \begin{align}
    \mat{K}^i{}_j (2\pi n/\omega, 0) &= \delta^i{}_j + 2\pi n \epsilon \left[\sqrt{1 - a^2} \pb{+} \mat{e}^i{}_j + \cos(\phi) \pb{\times} \mat{e}^i{}_j\right] \nonumber \\
    &\hspace{1em}- \frac{\pi n \epsilon^2}{2} \left\{\left[2 \pi n + 3 \sin(2 \phi) a^2\right] \delta^i{}_j - 12 \sin(\phi) a \sqrt{1 - a^2} \mat{\epsilon}^i{}_j\right\} + O(\epsilon^3), \\
    \mat{H}^i{}_j (2\pi n/\omega, 0) &= \delta^i{}_j - 2 \epsilon \sin(\phi) a \pb{\times} \mat{e}^i{}_j \nonumber \\
    &\hspace{1em}- \frac{\epsilon^2}{2} \left[\frac{4 \pi^2 n^2 + 9 [\cos(2 \phi) - 1] - 15}{6} \delta^i{}_j - 4 \pi \sin(\phi) a \sqrt{1 - a^2} n \mat{\epsilon}^i{}_j\right] + O(\epsilon^3), \displaybreak[0] \\
    \left.\partial_u \mat{K}^i{}_j (u, 0)\right|_{u = 2\pi n/\omega} &= -\omega \pi n \epsilon^2 \{[\cos (2\phi) - 1] a^2 + 3\} \delta^i{}_j + O(\epsilon^3), \label{eqn:dA_model} \\
    \left.\partial_u \mat{H}^i{}_j (u, 0)\right|_{u = 2\pi n/\omega} &= -\omega \epsilon \left[\sqrt{1 - a^2} \pb{+} \mat{e}^i{}_j + a \frac{\pi n \cos(\phi) - \sin(\phi)}{\pi n} \pb{\times} \mat{e}^i{}_j\right] \nonumber \\
    &\hspace{1em}-\frac{\omega \epsilon^2}{2} \left[\frac{8 \pi^2 n^2 - 9 a^2 [2 \pi \sin(2 \phi) n - \cos (2 \phi) + 1] + 15}{12 \pi n} \delta^i{}_j - 4 \sin(\phi) a \sqrt{1 - a^2} \mat{\epsilon}^i{}_j\right] + O(\epsilon^3).
  \end{align}
\end{subequations}
\end{widetext}
The terms in these expressions that are proportional to $\pb{+} \mat{e}^i{}_j$ and $\pb{\times} \mat{e}^i{}_j$ are the symmetric trace-free pieces, and as remarked only occur at first order.
As expected, Eq.~\eqref{eqn:dA_model} implies that the ``velocity memory'' of this waveform vanishes at first order.
At second order, there are pure trace pieces proportional to $\delta^i{}_j$ and antisymmetric pieces proportional to $\epsilon^i{}_j$.
As above, the antisymmetric pieces only occur when the polarization is not linear ($\phi \neq 0$).
To study the long-time behavior of these solutions, consider the regime where $n \to \infty$ as $\epsilon \to 0$, with

\begin{equation}
  n \sim \frac{1}{\epsilon^{1 - \eta}}.
\end{equation}
We assume $0 < \eta < 1$ so that the series~\eqref{eqn:2d_jacobi_expansion} converges.
In this regime, the antisymmetric pieces in Eqs.~\eqref{eqn:2d_jacobi_model} are subleading compared to the symmetric pieces.

\section{Persistent Observables} \label{sec:observables}

In this section, we review the persistent observables that we discussed in our first paper: the curve deviation, holonomy, and spinning test particle observables~\cite{Flanagan2019}.
For the first two of these observables, we can use the fact that the geodesic equation has exact solutions in plane wave spacetimes in terms of the transverse Jacobi propagators, as reviewed in Sec.~\ref{sec:bitensors}.
This allows us to find expressions which are nonperturbative in the initial separation and relative velocity.
For observables whose definitions involve accelerated curves, see Appendix~\ref{app:acceleration}; the results in these cases are perturbative in the acceleration.

The spinning test particle observable, however, does not have such a nonperturbative treatment, and so we use the results of~\cite{Flanagan2019} that are perturbative in separation, specialized to the class of plane wave spacetimes.
We could have used the same technique to derive perturbative results for the first two observables in plane wave spacetimes, but we did not because we already have analytic, nonperturbative results.

\begin{figure}[t]
  \includegraphics{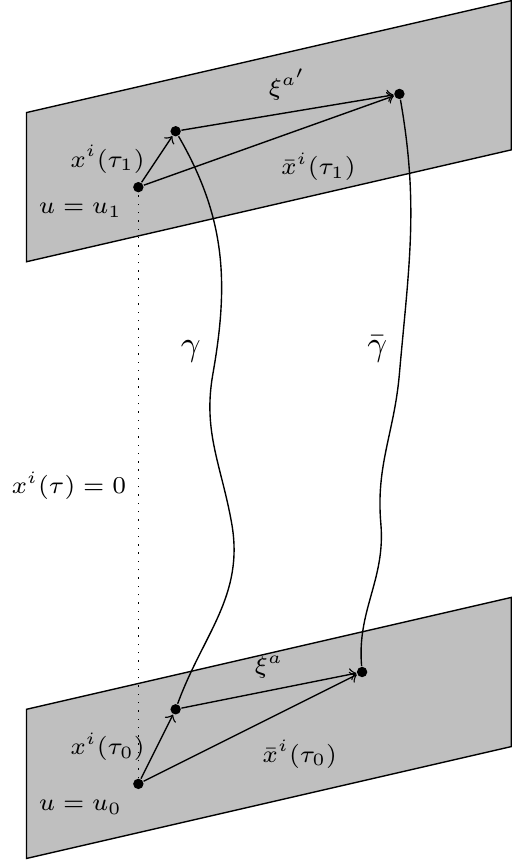}
  \caption{\label{fig:setup} The common setup for all persistent observables discussed in this paper: two timelike curves $\gamma$ and $\bar{\gamma}$ that have some initial separation $\xi^a$ at time $\tau_0$ and final separation $\xi^{a'}$ at time $\tau_1$.
    Intersecting these two curves are two planes of constant $u$ (the $v$ coordinate in this diagram is suppressed).
    The $x^i$ coordinates of the points $\gamma(\tau_0)$, $\gamma(\tau_1)$, $\bar{\gamma} (\tau_0)$, and $\bar{\gamma} (\tau_1)$ are also shown in this diagram.}
\end{figure}

We now introduce two pieces of notation that are used extensively in this section.
First, as persistent observables are defined with respect to an interval of proper time, we denote the initial time by $\tau_0$ and the final time by $\tau_1$; intermediate times are denoted by $\tau_2$, $\tau_3$, etc.
For a curve $\gamma$, points $\gamma(\tau_n)$ are denoted by $x$, $x'$, $x''$, etc., where $n$ is the number of primes, so $x \equiv \gamma(\tau_0)$, $x' \equiv \gamma(\tau_1)$, $x'' \equiv \gamma(\tau_2)$, etc.
The coordinates of these points are given by $u_n$, $v_n$, and $x^i (\tau_n)$.
This convention also holds for curves denoted by $\gamma$ with some sort of diacritical marking above or below: we apply the same diacritical mark to the point in question as well [e.g., $\bar{x}'$ refers to $\bar{\gamma} (\tau_1)$ and has coordinates $\bar{u}_1$, $\bar{v}_1$, and $\bar{x}^i (\tau_1)$].
A figure showing the setup common to all persistent observables discussed in this paper is given in Fig.~\ref{fig:setup}.

Second, many of the results in this section depend not only upon $\mat{\mathcal A}_{ij} (u)$ and the propagators $\mat{K}^i{}_j (u', u)$ and $\mat{H}^i{}_j (u', u)$, which are only functions of $u$ and $u'$, but also upon $x^i (\tau_0)$ and $\dot{x}^i (\tau_0)$.
This dependence is at most polynomial for the observables which we consider.
For some bitensor component $Q^{\cdots}{}_{\cdots}$ in Brinkmann coordinates (for simplicity we suppress the indices) that depends on $x^i (\tau_0)$ and $\dot{x}^i (\tau_0)$, we can write

\begin{equation} \label{eqn:polynomial_decomp}
  \begin{split}
    Q^{\cdots}{}_{\cdots} \equiv \sum_{k, m} \pb{x^k \dot{x}^m} \left[Q^{\cdots}{}_{\cdots}\right]_{i_1 \cdots i_k j_1 \cdots j_m} &x^{i_1} (\tau_0) \cdots x^{i_k} (\tau_0) \\
    &\times \dot{x}^{j_1} (\tau_0) \cdots \dot{x}^{j_m} (\tau_0).
  \end{split}
\end{equation}
Examples of this notation occur throughout this section; for example, in Eqs.~\eqref{eqn:curve_dev_results_K_x} and~\eqref{eqn:curve_dev_results_K_dotx} the quantities $\pb{x} [\Delta K^{v'}{}_i]_j (\tau_0)$ and $\pb{\dot x} [\Delta K^{v'}{}_i]_j (\tau_0)$ are coefficients in the expansion of the component $\Delta K^{v'}{}_i (\tau_0)$ of $\Delta K^{a'}{}_b$ in powers of $x^i (\tau_0)$ and $\dot{x}^i (\tau_0)$:

\begin{equation}
  \begin{split}
    \Delta K^{v'}{}_i (\tau_0) &= \pb{x} [\Delta K^{v'}{}_i]_j (\tau_0) x^j (\tau_0) \\
    &\hspace{1em}+ \pb{\dot x} [\Delta K^{v'}{}_i]_j (\tau_0) \dot{x}^j (\tau_0).
  \end{split}
\end{equation}
There are often relationships between the coefficients that occur in these expansions; see for example Eq.~\eqref{eqn:spinning_results}.

\subsection{Curve deviation observable} \label{sec:curve_dev}

The curve deviation observable is defined as follows~\cite{Flanagan2019}.
Consider two observers following timelike curves $\gamma$ and $\bar{\gamma}$.
At two points $x$ and $\bar{x}$ along $\gamma$ and $\bar{\gamma}$ (respectively), the two observers synchronize their clocks such that $x = \gamma(\tau_0)$ and $\bar{x} = \bar{\gamma} (\tau_0)$.
At $\gamma(\tau_0)$, the observer following $\gamma$ measures the initial separation $\xi^a$ (as defined using the exponential map) and relative velocity $\dot{\xi}^a \equiv \uD \xi^a/\ud \tau$ between $\bar{\gamma}$ and $\gamma$.
Moreover, the two observers monitor their accelerations until some proper time $\tau_1$, when the observer following $\gamma$ measures the separation $\xi^{a'}$ between $\gamma(\tau_1)$ and $\bar{\gamma} (\tau_1)$.
This definition of final separation is known as the \emph{isochronous correspondence}~\cite{Vines2014a}.
The observers then compare their final separation to a prediction based on their previous measurements that would be correct had they been in a flat region of spacetime during the interval from $\tau_0$ to $\tau_1$.
There are many possible predictions that an observer could make, but a straightforward covariant prediction is the one given by parallel transport.
The difference between the final separation and the prediction, which we call $\Delta \xi^{a'}_{\textrm{CD}}$, is our curve deviation observable:

\begin{widetext}
\begin{equation} \label{eqn:curve_deviation_def}
  \Delta \xi^{a'}_{\textrm{CD}} \equiv \xi^{a'} - \pb{\gamma} g^{a'}{}_a \left[\xi^a + (\tau_1 - \tau_0) \dot{\xi}^a\right] - \int_{\tau_0}^{\tau_1} \ud \tau_2 \int_{\tau_0}^{\tau_2} \ud \tau_3 \pb{\gamma} g^{a'}{}_{a'''} \left(g^{a'''}{}_{\bar a'''} \ddot{\bar \gamma}^{\bar a'''} - \ddot{\gamma}^{a'''}\right).
\end{equation}
\end{widetext}

In our previous paper, we showed that $\Delta \xi^{a'}_{\textrm{CD}}$ could be written perturbatively in the initial separation, initial relative velocity, and accelerations of $\gamma$ and $\bar{\gamma}$.
For convenience, we restrict here to the case where $\gamma$ and $\bar{\gamma}$ are geodesic.
The general case is treated in Appendix~\ref{app:acceleration}.
The particular expansion was of the form

\begin{equation}
  \begin{split} \label{eqn:curve_dev_decomp}
    \Delta \xi^{a'}_{\textrm{CD}} \equiv &\left[\Delta K^{a'}{}_b + L^{a'}{}_{bc} \xi^c + N^{a'}{}_{bc} \dot{\xi}^c\right] \xi^b \\
    &+ \left[(\tau_1 - \tau_0) \Delta H^{a'}{}_b + M^{a'}{}_{bc} \dot{\xi}^c\right] \dot{\xi}^b \\
    &+ O(\boldsymbol{\xi}, \dot{\boldsymbol{\xi}})^3, \\
  \end{split}
\end{equation}
which serves as a definition of the bitensors on the right-hand side.
The bitensors $\Delta K^{a'}{}_a$ and $\Delta H^{a'}{}_a$ are closely related to the Jacobi propagators, as they are given by $\Delta K^{a'}{}_a = \pb{\gamma} K^{a'}{}_a - \pb{\gamma} g^{a'}{}_a$ and $\Delta H^{a'}{}_a = \pb{\gamma} H^{a'}{}_a - \pb{\gamma} g^{a'}{}_a$.

In plane wave spacetimes, the geodesic equation has exact solutions in terms of transverse Jacobi propagators.
As such, curve deviation, at least restricting to the case where there is no acceleration for either curve, will have an exact solution, instead of a perturbative solution in the separation of the two particles.

In order to compute the observables~\eqref{eqn:curve_dev_decomp}, we need the separation vector in the flat regions of the plane wave spacetime in Brinkmann coordinates.
Before the burst, this separation vector is given by

\begin{equation} \label{eqn:xi_exact}
  \xi^a = (\bar{u}_0 - u_0) (\partial_u)^a + [\bar{x}^i (\tau_0) - x^i (\tau_0)] (\partial_i)^a - (\bar{v}_0 - v_0) \ell^a.
\end{equation}
A similar expression holds for $\xi^{a'}$, the separation vector after the burst.
Another piece that is required for the curve deviation observable is the relative velocity.
Here we use the fact that, in the flat regions, $\dot{\xi}^a = g^a{}_{\bar a} \dot{\bar \gamma}^a - \dot{\gamma}^a$;  this implies [from Eq.~\eqref{eqn:four_velocity}] that

\begin{equation} \label{eqn:dotxi_exact}
  \begin{split}
    \dot{\xi}^a = (\bar{\chi} - \chi) (\partial_u)^a &+ \left[\dot{\bar x}^i (\tau_0) - \dot{x}^i (\tau_0)\right] (\partial_i)^a \\
    &- \bigg\{\frac{1}{2 \bar \chi} \left[1 + \dot{\bar x}^i (\tau_0) \dot{\bar x}_i (\tau_0)\right] \\
    &\hspace{2em}- \frac{1}{2 \chi} \left[1 + \dot{x}^i (\tau_0) \dot{x}_i (\tau_0)\right]\bigg\} \ell^a,
  \end{split}
\end{equation}
where $\bar{\chi} \equiv \dot{\bar \gamma}^a \ell_a$.

At this point, we note that this calculation is greatly simplified in the case where we assume that $\bar{u}_0 = u_0$ (which implies that $\xi^a \ell_a = 0$) and $\bar{\chi} = \chi$ (which implies that $\dot{\xi}^a \ell_a = 0$).
In particular, this assumption about the initial data means that the exact solutions are \emph{quadratic} in $\xi^a$ and $\dot{\xi}^a$; in general, the results are not polynomial in $\xi^a \ell_a$ and $\dot{\xi}^ a \ell_a$.
Note that this assumption implies that $\bar{u}_1 = u_1$ as well, from Eq.~\eqref{eqn:u_geo}.
Thus, we are also associating points on the two worldlines with equal values of $u$, the gravitational wave phase, so this restriction could be called the \emph{isophase correspondence}.

Taking into account these assumptions, we find that $\xi^{a'}$ is given by

\begin{equation}
  \xi^{a'} = \left[\bar{x}^i (\tau_1) - x^i (\tau_1)\right] (\partial_i)^{a'} - (\bar{v}_1 - v_1) \ell^{a'}.
\end{equation}
We can determine the first term in this equation by using Eq.~\eqref{eqn:x_geo}, together with Eqs.~\eqref{eqn:xi_exact} and~\eqref{eqn:dotxi_exact}:

\begin{equation} \label{eqn:xi_soln}
  \begin{split}
    \bar{x}^i (\tau_1) - x^i (\tau_1) &= \mat{K}^i{}_j (u_1, u_0) \xi^j (\tau_0) \\
    &\hspace{1em}+ (\tau_1 - \tau_0) \mat{H}^i{}_j (u_1, u_0) \dot{\xi}^j (\tau_0).
  \end{split}
\end{equation}
For the second term, we use Eqs.~\eqref{eqn:v_geo} and~\eqref{eqn:dotxi_exact}:

\begin{widetext}
\begin{equation}
  \begin{split}
    \bar{v}_1 - v_1 = \bar{v}_0 - v_0 &+ \frac{1}{2\chi} \left\{\left[\bar{x}^i (\tau_1) \dot{\bar x}_i (\tau_1) - x^i (\tau_1) \dot{x}_i (\tau_1)\right] - \left[\bar{x}^i (\tau_0) \dot{\bar x}_i (\tau_0) - x^i (\tau_0) \dot{x}_i (\tau_0)\right]\right\} \\
    = \bar{v}_0 - v_0 &+ \mat{K}^k{}_i (u_1, u_0) \partial_{u_1} \mat{K}{}_{kj} (u_1, u_0) \left[\frac{1}{2} \xi^i (\tau_0) \xi^j (\tau_0) + \xi^i (\tau_0) x^j (\tau_0)\right] \\
    &+ \frac{1}{\chi} \left\{\mat{K}^k{}_i (u_1, u_0) \partial_{u_1} \left[(u_1 - u_0) \mat{H}{}_{kj} (u_1, u_0)\right] - \delta_{ij}\right\} \left[\frac{1}{2} \xi^i (\tau_0) \dot{\xi}^j (\tau_0) + \xi^i (\tau_0) \dot{x}^j (\tau_0)\right] \\
    &+ \frac{1}{\chi} (u_1 - u_0) \mat{H}^k{}_i (u_1, u_0) \partial_{u_1} \mat{K}{}_{kj} (u_1, u_0) \left[\frac{1}{2} \dot{\xi}^i (\tau_0) \xi^j (\tau_0) + \dot{\xi}^i (\tau_0) x^j (\tau_0)\right] \\
    &+ \frac{1}{\chi^2} (u_1 - u_0) \mat{H}^k{}_i (u_1, u_0) \partial_{u_1} \left[(u_1 - u_0) \mat{H}{}_{kj} (u_1, u_0)\right] \left[\frac{1}{2} \dot{\xi}^i (\tau_0) \dot{\xi}^j (\tau_0) + \dot{\xi}^i (\tau_0) \dot{x}^j (\tau_0)\right].
  \end{split}
\end{equation}

At this point, note that the curve deviation observable is defined as the result of geodesic deviation with the prediction in flat spacetime subtracted off.
This prediction is given by

\begin{equation}
  \begin{split}
    \pb{\gamma} g^{a'}{}_a \left[\xi^a + (\tau_1 - \tau_0) \dot{\xi}^a\right] &= \left[\xi^i (\tau_0) + (\tau_1 - \tau_0) \dot{\xi}^i (\tau_0)\right] \bigg\{(\partial_i)^{a'} - \bigg[\partial_{u_1} \mat{K}{}_{ij} (u_1, u_0) x^j (\tau_0) \\
    &\hspace{17.5em}+ \frac{1}{\chi} \left\{\partial_{u_1} \left[(u_1 - u_0) \mat{H}{}_{ij} (u_1, u_0)\right] - \delta_{ij}\right\} \dot{x}^j (\tau_0)\bigg] \ell^{a'}\bigg\} \\
    &\hspace{1em}- \left\{\bar{v}_0 - v_0 + \frac{1}{\chi^2} (u_1 - u_0) \left[\frac{1}{2} \dot{\xi}^i (\tau_0) \dot{\xi}_i (\tau_0) + \dot{\xi}^i (\tau_0) \dot{x}_i (\tau_0)\right]\right\} \ell^{a'},
  \end{split}
\end{equation}
where we have used Eq.~\eqref{eqn:4d_propagators} to compute the parallel propagator.
We now use the decompositions in Eqs.~\eqref{eqn:curve_dev_decomp} and~\eqref{eqn:polynomial_decomp}.
Since the exact solutions are quadratic in $\xi^a$ and $\dot\xi^a$, we just need to write down the components of $\Delta K^{a'}{}_a$, $\Delta H^{a'}{}_a$, $L^{a'}{}_{bc}$, $N^{a'}{}_{bc}$, and $M^{a'}{}_{bc}$.
The nonvanishing quantities needed to compute these bitensors (and thus, the curve deviation observable) are as follows:

\begin{subequations} \label{eqn:curve_dev_results}
  \begin{align}
    \Delta K^i{}_j (\tau_1, \tau_0) &= \mat{K}^i{}_j (u_1, u_0) - \delta^i{}_j, \\
    \pb{x} [\Delta K^{v'}{}_i]_j (\tau_0) &= \partial_{u_1} \mat{K}{}_{kj} (u_1, u_0) \Delta K^k{}_i (\tau_1, \tau_0), \label{eqn:curve_dev_results_K_x} \\
    \pb{\dot x} [\Delta K^{v'}{}_i]_j (\tau_0) &= \frac{1}{\chi} \partial_{u_1} \left[(u_1 - u_0) \mat{H}{}_{kj} (u_1, u_0)\right] \Delta K^k{}_i (\tau_0), \label{eqn:curve_dev_results_K_dotx} \\
    \Delta H^i{}_j (\tau_1, \tau_0) &= \mat{H}^i{}_j (u_1, u_0) - \delta^i{}_j, \\
    \pb{x} [\Delta H^{v'}{}_i]_j (\tau_0) &= \partial_{u_1} \mat{K}{}_{kj} (u_1, u_0) \Delta H^k{}_i (\tau_1, \tau_0), \\
    \pb{\dot x} [\Delta H^{v'}{}_i]_j (\tau_0) &= \frac{1}{\chi} \partial_{u_1} \left[(u_1 - u_0) \mat{H}{}_{kj} (u_1, u_0)\right] \Delta H^k{}_i (\tau_0), \\
    L^{v'}{}_{ij} (\tau_0) &= \frac{1}{2} \mat{K}{}_{k(i} (u_1, u_0) \partial_{u_1} \mat{K}^k{}_{j)} (u_1, u_0), \label{eqn:curve_dev_results_L} \\
    N^{v'}{}_{ij} (\tau_0) &= \frac{1}{2\chi} \left\{\partial_{u_1} \left[(u_1 - u_0) \mat{K}^k{}_i (u_1, u_0) \mat{H}{}_{kj} (u_1, u_0)\right] - \delta_{ij}\right\}, \\
    M^{v'}{}_{ij} (\tau_0) &= \frac{1}{2\chi^2} (u_1 - u_0) \left\{\mat{H}{}_{k(i} (u_1, u_0) \partial_{u_1} \left[(u_1 - u_0) \mat{H}^k{}_{j)} (u_1, u_0)\right] - \delta_{ij}\right\}.
  \end{align}
\end{subequations}
\end{widetext}
The result~\eqref{eqn:curve_dev_results} makes it clear that the curve deviation observable depends only on the transverse Jacobi propagators and their first derivatives at $\tau_1$, with no need to integrate any additional quantities from $\tau_0$ to $\tau_1$.

Moreover, note that $L^i{}_{jk} (\tau_1, \tau_0)$, $N^i{}_{jk} (\tau_1, \tau_0)$, and $M^i{}_{jk} (\tau_1, \tau_0)$ all vanish.
This is a consequence of the fact that geodesic deviation, in the case where the initial separation lies entirely in a surface of constant $u$ and $v$, has no corrections at second order in the separation, at least for the components that also lie in this surface.
Because of this property, the proper time delay observable $\dot{\gamma}_{a'} \Delta \xi^{a'}_{\textrm{CD}}$ (described in greater detail in~\cite{Flanagan2019}) can be expressed as

\begin{equation}
  \begin{split}
    \dot{\gamma}_{a'} \Delta \xi^{a'}_{\textrm{CD}} = -\chi \Big[&L^{v'}{}_{ij} (\tau_0) \xi^i (\tau_0) \xi^j (\tau_0) \\
    &+ N^{v'}{}_{ij} (\tau_0) \xi^i (\tau_0) \dot{\xi}^j (\tau_0) \\
    &+ M^{v'}{}_{ij} (\tau_0) \dot{\xi}^i (\tau_0) \dot{\xi}^j (\tau_0)\Big].
  \end{split}
\end{equation}

\subsection{Holonomies} \label{sec:holonomy}

The definition of our holonomy observable is as follows~\cite{Flanagan2019}: consider a closed loop that is composed of segments of two timelike curves $\gamma$ and $\bar{\gamma}$ that are connected by two spacelike geodesics.
These spacelike geodesics intersect $\gamma$ and $\bar{\gamma}$ at values of proper time along each curve given by $\tau_0$ and $\tau_1$.
Consider initial data $P^a$ and $J^{ab}$ at $\gamma(\tau_0)$, where $J^{ab}$ is antisymmetric.
These initial data represent linear and angular momenta that have been measured by an observer.
For example, they could be the linear and angular momenta either of a point particle or of the spacetime itself.
Starting with these initial data, the observers then transport the linear and angular momenta around the closed loop mentioned above according to the following differential equation:

\begin{subequations} \label{eqn:transport}
  \begin{align}
    k^b \nabla_b P^a &= -\vkappa{K}{}^a{}_{bcd} k^b J^{cd}, \\
    k^c \nabla_c J^{ab} &= 2 P^{[a} k^{b]}.
  \end{align}
\end{subequations}
Here $k^a$ is the tangent to the loop and $\vkappa{K}{}^a{}_{bcd}$ is a tensor constructed from the Riemann tensor (and the metric) that depends on a set of four constant parameters $\varkappa = (\varkappa_1, \varkappa_2, \varkappa_3, \varkappa_4)$:

\begin{equation} \label{eqn:K_def}
  \begin{split}
    \vkappa{K}{}^{ab}{}_{cd} = \varkappa_1 R^{ab}{}_{cd} &+ \varkappa_2 \delta^a{}_{[c} R^b{}_{d]} + \varkappa_3 \delta^b{}_{[c} R^a{}_{d]} \\
    &+ \varkappa_4 R \delta^a{}_{[c} \delta^b{}_{d]}.
  \end{split}
\end{equation}
In this paper, we restrict to $\varkappa_1$ being the only nonvanishing component of $\varkappa$, and we further restrict the value of $\varkappa_1$ to be either $0$ or $1/2$.
Equation~\eqref{eqn:transport} is solved first along $\gamma$, then along the geodesic between $\gamma(\tau_1)$ and $\bar{\gamma} (\tau_1)$, then along $\bar{\gamma}$ (in the reverse direction), and then finally along the geodesic between $\bar{\gamma} (\tau_0)$ and $\gamma(\tau_0)$.

The final values obtained by solving the differential equations in Eq.~\eqref{eqn:transport} we will denote by $\vkappa{P}{}^a$ and $\vkappa{J}{}^{ab}$.
As our differential equations are linear, these final values depend linearly on the initial data $P^a$ and $J^{ab}$; the matrix which describes the linear relationship is our holonomy observable:

\begin{subequations} \label{eqn:holonomy_matrix}
  \begin{align}
    \vkappa{P}{}^a &\equiv \smallunderset{PP}{\vkappa \Lambda}{}^a{}_c (\gamma, \bar{\gamma}; \tau_1) P^c + \smallunderset{PJ}{\vkappa \Lambda}{}^a{}_{cd} (\gamma, \bar{\gamma}; \tau_1) J^{cd}, \\
    \vkappa{J}{}^{ab} &\equiv \smallunderset{JP}{\vkappa \Lambda}{}^{ab}{}_c (\gamma, \bar{\gamma}; \tau_1) P^c + \smallunderset{PP}{\vkappa \Lambda}{}^{ab}{}_{cd} (\gamma, \bar{\gamma}; \tau_1) J^{cd}.
  \end{align}
\end{subequations}
The motivation for this definition was discussed in~\cite{Flanagan2014, Flanagan2016, Flanagan2019}; in particular, this holonomy contains the displacement, relative velocity, relative proper time, and relative rotation persistent observables for $\varkappa = (0,0,0,0)$.

For convenience, we introduce notation used in~\cite{Flanagan2019, Flanagan2016}, where the combination of $P^a$ and $J^{ab}$ by a single vector $X^A$ was denoted by:

\begin{equation} \label{eqn:X_def}
  X^A \equiv \begin{pmatrix}
    P^a \\
    J^{ab}
  \end{pmatrix}.
\end{equation}
In this notation, Eq.~\eqref{eqn:holonomy_matrix} becomes

\begin{equation} \label{eqn:Lambda_def}
  \vkappa{X}{}^A = \vkappa{\Lambda}{}^A{}_B (\gamma, \bar{\gamma}; \tau_1) X^B.
\end{equation}
We use the same notation that we used for the components of $\vkappa{\Lambda}{}^A{}_B (\gamma, \bar{\gamma}; \tau_1)$ for other matrices that act on the space of linear and angular momentum:

\begin{equation} \label{eqn:mat_def}
  A^A{}_C \equiv \begin{pmatrix}
    \smallunderset{PP}{A}^a{}_c & \smallunderset{PJ}{A}^a{}_{cd} \\
    \smallunderset{JP}{A}^{ab}{}_c & \smallunderset{JJ}{A}^{ab}{}_{cd}
  \end{pmatrix}.
\end{equation}

Note that the holonomy observable is \emph{nonzero} in flat space, even though it is trivial, being given by $\delta^A{}_B$.
As such, we find it convenient to define

\begin{equation} \label{eqn:Omega_def}
  \vkappa{\Omega}{}^A{}_B (\gamma, \bar{\gamma}; \tau_1) \equiv \vkappa{\Lambda}{}^A{}_B (\gamma, \bar{\gamma}; \tau_1) - \delta^A{}_B,
\end{equation}
as the persistent observable associated with the holonomy.

In general spacetimes, we needed to expand perturbatively in the separation and relative velocity of the two curves $\gamma$ and $\bar{\gamma}$; for plane waves, in contrast, these calculations can be done nonperturbatively.
We perform only the nonperturbative calculations in this section.
Note that we also continue to use the assumption that $\xi^a \ell_a = 0$ and $\dot{\xi}^a \ell_a = 0$, for simplicity.

We consider the holonomies for the two types of transport considered in~\cite{Flanagan2019}: $\varkappa = (0, 0, 0, 0)$ (affine transport) and $\varkappa = (1/2, 0, 0, 0)$ (dual Killing transport).
The affine transport holonomy can be recast as a Poincar\'e transformation, as it can be written in terms of a vector $\Delta \chi^a (\gamma, \bar{\gamma}; \tau_1)$ and the holonomy of the metric-compatible connection (which is itself a Lorentz transformation)~\cite{Flanagan2014}.
On the other hand, the holonomy of dual Killing transport does not share this property.
Instead, it can be thought of as the final linear and angular momentum that would arise from using the Mathisson-Papapetrou equations to transport these momenta around a closed curve.
The holonomy of affine transport we denote by $\zero{\Lambda}{}^A{}_B (\gamma, \bar{\gamma}; \tau_1)$, with the ``$0$'' indicating that $\varkappa = (0, 0, 0, 0)$.\footnote{Note that this is in contrast to the notation used in~\cite{Flanagan2019}, where we denoted this holonomy by $\mathring{\Lambda}{}^A{}_B (\gamma, \bar{\gamma}; \tau_1)$.}
The holonomy in the case of dual Killing transport we denote by $\half{\Lambda}{}^A{}_B (\gamma, \bar{\gamma}; \tau_1)$, with the ``$1/2$'' indicating that $\varkappa = (1/2, 0, 0, 0)$.

\subsubsection{Affine transport}

First, we consider the case of affine transport.
Here we take advantage of the fact that~\cite{Flanagan2014, Flanagan2019}

\begin{subequations} \label{eqn:0_holonomy_assembly}
  \begin{align}
    \smallunderset{PP}{\zero \Lambda}{}^a{}_c (\gamma, \bar{\gamma}; \tau_1) &= \Lambda^a{}_c (\gamma, \bar{\gamma}; \tau_1), \label{eqn:0_holonomy_pp} \\
    \smallunderset{JP}{\zero \Lambda}{}^{ab}{}_c (\gamma, \bar{\gamma}; \tau_1) &= 2 \Delta \chi^{[a} (\gamma, \bar{\gamma}; \tau_1) \Lambda^{b]}{}_c (\gamma, \bar{\gamma}; \tau_1), \label{eqn:0_holonomy_jp} \\
    \smallunderset{JJ}{\zero \Lambda}{}^{ab}{}_{cd} (\gamma, \bar{\gamma}; \tau_1) &= \Lambda^{[a}{}_c (\gamma, \bar{\gamma}; \tau_1) \Lambda^{b]}{}_d (\gamma, \bar{\gamma}; \tau_1),
  \end{align}
\end{subequations}
where $\Lambda^a{}_b (\gamma, \bar{\gamma}; \tau_1)$ is the holonomy of parallel transport with respect to the usual metric-compatible connection and $\Delta \chi^a (\gamma, \bar{\gamma}; \tau_1)$ is given by

\vspace{2em}
\begin{equation} \label{eqn:Delta_chi}
  \begin{split}
    \Delta &\chi^a (\gamma, \bar{\gamma}; \tau_1) \\
    &= \xi^a - \Lambda^a{}_b (\gamma, \bar{\gamma}; \tau_1) \pb{\gamma} g^b{}_{b'} \left[\xi^{b'} - (\tau_1 - \tau_0) \dot{\xi}^{b'}\right].
  \end{split}
\end{equation}
Therefore, all we need in order to solve for the holonomy of affine transport is the value of the separation $\xi^{a'}$ at $\tau_1$ and the holonomy $\Lambda^a{}_b (\gamma, \bar{\gamma}; \tau_1)$ of the metric-compatible connection around the loop.

We computed the separation $\xi^{a'}$ in Sec.~\ref{sec:curve_dev}, so at this point we merely need to compute $\Lambda^a{}_b (\gamma, \bar{\gamma}; \tau_1)$; for simplicity, we define

\begin{equation} \label{eqn:parallel_Omega}
  \Omega^a{}_b (\gamma, \bar{\gamma}; \tau_1) \equiv \Lambda^a{}_b (\gamma, \bar{\gamma}; \tau_1) - \delta^a{}_b
\end{equation}
[in analogy to Eq.~\eqref{eqn:Omega_def}].
Since $\ell^a$ is covariantly constant, it follows that
$\Omega^a{}_b (\gamma, \bar{\gamma}; \tau_1) \ell^b = 0$.
After a lengthy calculation using our expressions for the parallel propagators in Eq.~\eqref{eqn:4d_propagators}, we find that

\begin{subequations} \label{eqn:holonomy_xi}
  \begin{align}
    \Omega^a{}_b &(\gamma, \bar{\gamma}; \tau_1) (\partial_i)^b \nonumber \\
    &= \frac{1}{\chi} \left[\dot{\xi}_i (\tau_1) - \dot{\xi}_i (\tau_0)\right] \ell^a, \\
    \Omega^a{}_b &(\gamma, \bar{\gamma}; \tau_1) u^b \nonumber \\
    &= -\frac{1}{\chi} \left[\dot{\xi}^i (\tau_1) - \dot{\xi}^i (\tau_0)\right] (\partial_i)^a \nonumber \\
    &\hspace{1em}- \frac{1}{2\chi^2} \left[\dot{\xi}^i (\tau_1) - \dot{\xi}^i (\tau_0)\right] \left[\dot{\xi}_i (\tau_1) - \dot{\xi}_i (\tau_0)\right] \ell^a.
  \end{align}
\end{subequations}
By a lengthy calculation involving Eq.~\eqref{eqn:Delta_chi} we can also show that

\begin{widetext}
\begin{subequations} \label{eqn:chi_xi}
  \begin{align}
    \Delta \chi^i (\gamma, \bar{\gamma}; \tau_1, \tau_0) &= \xi^i (\tau_0) - \xi^i (\tau_1) + (\tau_1 - \tau_0) \dot{\xi}^i (\tau_1), \\
    \Delta \chi^v (\gamma, \bar{\gamma}; \tau_1) &= \frac{1}{\chi} \bigg\{\left[\frac{1}{2} \dot{\xi}_i (\tau_1) - \dot{\xi}_i (\tau_0)\right] \left[\xi^i (\tau_1) - (\tau_1 - \tau_0) \dot{\xi}^i (\tau_1)\right] + \frac{1}{2} \xi^i (\tau_0) \dot{\xi}_i (\tau_0)\bigg\} + \frac{1}{\chi} \dot{x}_i (\tau_0) \Delta \chi^i (\tau_1, \tau_0).
  \end{align}
\end{subequations}
\end{widetext}

Equations~\eqref{eqn:holonomy_xi} and~\eqref{eqn:chi_xi} can be used to determine the nonzero components of $\zero{\Omega}{}^A{}_B (\gamma, \bar{\gamma}; \tau_1)$, and then to find the values of these components in plane wave spacetimes as a function of initial data $x^i (\tau_0)$, $\xi^i (\tau_0)$, and $\dot{\xi}^i (\tau_0)$.
These results are given in Appendix~\ref{app:holonomy}, in Eqs.~\eqref{eqn:affine_comps},~\eqref{eqn:affine_expansion}, and~\eqref{eqn:affine_coeffs}.

\subsubsection{Dual Killing transport}

We can also show that the holonomy of dual Killing transport can be written in terms of the holonomy of affine transport, just as that holonomy can be written in terms of the holonomy of the metric-compatible connection.
To show this, we note that because the beginning and end of the loop are in the flat regions of spacetime, in this region there is no difference between affine transport and dual Killing transport (the value of $\varkappa$ is irrelevant, as the Riemann tensor vanishes).
Therefore, we can compute the holonomy by using different values of $\varkappa$ along different segments of the loop.
This yields

\begin{equation} \label{eqn:dual_killing_affine_core}
  \begin{split}
    \half{\Omega}{}^A{}_B (\gamma, \bar{\gamma}; \tau_1) &= \Big\{\left[\delta^A{}_C + \hat{\Delta}^A{}_C (\gamma, \bar{\gamma}; \tau_1)\right] \zero{\Omega}{}^C{}_D (\gamma, \bar{\gamma}; \tau_1) \\
    &\hspace{1.5em}+ \hat{\Delta}^A{}_D (\gamma, \bar{\gamma}; \tau_1) - \hat{\Delta}^A{}_D (\gamma, \gamma; \tau_1)\Big\} \\
    &\hspace{1em}\times \left[\delta^D{}_B + \Delta^D{}_B (\gamma; \tau_1)\right],
  \end{split}
\end{equation}
where

\begin{subequations} \label{eqn:Delta_defs}
  \begin{align}
    \Delta^A{}_B (\gamma; \tau_1) &\equiv \pb{\gamma} \zero{g}{}^A{}_{A'} \pb{\gamma} \half{g}{}^{A'}{}_B - \delta^A{}_B, \\
    \hat{\Delta}^A{}_B (\gamma, \bar{\gamma}; \tau_1) &\equiv \zero{g}{}^A{}_{\bar A} \pb{\bar \gamma} \half{g}{}^{\bar A}{}_{\bar A'} \pb{\gamma} \zero{g}{}^{\bar A'}{}_{\bar B} \zero{g}{}^{\bar B}{}_B - \delta^A{}_B. \label{eqn:hat_Delta_def}
  \end{align}
\end{subequations}
Note that in Eq.~\eqref{eqn:dual_killing_affine_core}, both $\hat{\Delta}^A{}_B (\gamma, \bar{\gamma}; \tau_1)$ and $\hat{\Delta}^A{}_B (\gamma, \gamma; \tau_1)$ appear.
The latter is defined by Eq.~\eqref{eqn:hat_Delta_def}, but with $\bar{\gamma} = \gamma$, which implies that $\bar{x} = x$ and $\bar{x}' = x'$; equivalently, $\hat{\Delta}^A{}_B (\gamma, \gamma; \tau_1)$ is the same as $\Delta^A{}_B (\gamma; \tau_1)$, but with the order of the parallel propagators reversed.

Equation~\eqref{eqn:dual_killing_affine_core} implies that there is a portion of the holonomy of dual Killing transport that is the same as the holonomy for affine transport.
This portion has the interpretation of being a Poincar\'e transformation.
We give expressions for the components of the various pieces of Eq.~\eqref{eqn:dual_killing_affine_core} in Appendix~\ref{app:holonomy}.
The key point to take away is that all components of the tensors that occur can be written solely in terms of the transverse Jacobi propagators $\mat{K}^i{}_j (u', u)$ and $\mat{H}^i{}_j (u', u)$ and their derivatives.

We now discuss the number of independent nonzero components of the holonomy $\half{\Omega}{}^A{}_B (\gamma, \bar{\gamma}; \tau_1)$ in plane wave spacetimes.
The holonomy~\eqref{eqn:dual_killing_affine_core}, in general spacetimes, has potentially 100 different independent, nonzero components.
Because of the five-dimensional space of Killing vector fields in plane wave spacetimes, our final result should have fewer independent components.
The easiest way to see this is to note that, for a given Killing vector $\xi^a$, and for $P^a$ and $J^{ab}$ transported along a curve using dual Killing transport,

\begin{equation}
  Q \equiv P^a \xi_a + \frac{1}{2} J^{ab} \nabla_a \xi_b
\end{equation}
is a constant along the curve.
In particular, this means that

\begin{subequations} \label{eqn:dual_killing_eigenvectors}
  \begin{align}
    0 &= \smallunderset{PP}{\half \Omega}{}^c{}_a (\gamma, \bar{\gamma}; \tau_1) \xi_c + \frac{1}{2} \smallunderset{JP}{\half \Omega}{}^{cd}{}_a (\gamma, \bar{\gamma}; \tau_1) \nabla_c \xi_d, \\
    0 &= \smallunderset{PJ}{\half \Omega}{}^c{}_{ab} (\gamma, \bar{\gamma}; \tau_1) \xi_c + \frac{1}{2} \smallunderset{JJ}{\half \Omega}{}^{cd}{}_{ab} (\gamma, \bar{\gamma}; \tau_1) \nabla_c \xi_d.
  \end{align}
\end{subequations}
The five Killing vectors for which this equation holds are $\xi_a = \ell_a$
(which satisfies $\nabla_{[a} \xi_{b]} = 0$), and $\xi_a = \Xi_a$, where $\Xi_a$ is given by Eq.~\eqref{eqn:killing}, and thus satisfy

\begin{equation}
  \nabla_{[a} \xi_{b]} = 2 \dot{\underline \Xi}_i (u_0) \ell_{[a} (\ud x^i)_{b]}.
\end{equation}
Therefore, Eqs.~\eqref{eqn:dual_killing_eigenvectors} imply that

\begin{subequations} \label{eqn:dual_killing_constraints}
  \begin{align}
    \smallunderset{PP}{\half \Omega}{}^u{}_\mu (\gamma, \bar{\gamma}; \tau_1) &= 0, \quad &\smallunderset{PJ}{\half \Omega}{}^u{}_{\mu\nu} (\gamma, \bar{\gamma}; \tau_1) &= 0, \label{eqn:killing_l} \\
    \smallunderset{PP}{\half \Omega}{}^i{}_\mu (\gamma, \bar{\gamma}; \tau_1) &= 0, \quad &\smallunderset{PJ}{\half \Omega}{}^i{}_{\mu\nu} (\gamma, \bar{\gamma}; \tau_1) &= 0, \label{eqn:killing_Xi} \\
    \smallunderset{JP}{\half \Omega}{}^{ui}{}_\mu (\gamma, \bar{\gamma}; \tau_1) &= 0, \quad &\smallunderset{JJ}{\half \Omega}{}^{ui}{}_{\mu\nu} (\gamma, \bar{\gamma}; \tau_1) &= 0. \label{eqn:killing_dotXi}
  \end{align}
\end{subequations}
Here, Eq.~\eqref{eqn:killing_l} corresponds to $\xi^a = \ell^a$, while Eqs.~\eqref{eqn:killing_Xi} and~\eqref{eqn:killing_dotXi} correspond to $\xi^a = \Xi^a$, and are the constraints due to varying over the initial data $\Xi^i (u_0)$ and $\dot{\Xi}^i (u_0)$ in Eq.~\eqref{eqn:killing_ics}, respectively.
This reduces the number of possible independent components to 50.

Comparing this general result~\eqref{eqn:dual_killing_constraints} to our calculation, we first note that na\"ively, the multiplication of products in Eq.~\eqref{eqn:dual_killing_affine_core} gives nonzero values of $\smallunderset{PP}{\half \Omega}{}^i{}_u (\gamma, \bar{\gamma}; \tau_1)$ and $\smallunderset{JP}{\half \Omega}{}^{ui}{}_u (\gamma, \bar{\gamma}; \tau_1)$, but a careful inspection of these components shows that they are zero.
A sketch of how to show this goes as follows: first, use Eqs.~\eqref{eqn:holonomy_xi} and~\eqref{eqn:chi_xi} to write the components of the affine transport holonomy in terms of $\xi^i (\tau_0)$, $\xi^i (\tau_1)$, and their derivatives, and also write out explicit expressions for the components of $\Delta^A{}_B (\gamma, \bar{\gamma}; \tau_1) - \Delta^A{}_B (\gamma, \gamma; \tau_1)$ in terms of $\xi^i (\tau_0)$ and $\dot{\xi}^i (\tau_0)$, using the results of Appendix~\ref{app:holonomy}.
Next, use the identities satisfied by the transverse Jacobi propagators in Eqs.~\eqref{eqn:switch_identities} and~\eqref{eqn:deriv_identities}, and finally use Eq.~\eqref{eqn:xi_soln} and its derivative, but with $\tau_0$ and $\tau_1$ switched.
Since $\smallunderset{PP}{\half \Omega}{}^i{}_u (\gamma, \bar{\gamma}; \tau_1)$ and $\smallunderset{JP}{\half \Omega}{}^{ui}{}_u (\gamma, \bar{\gamma}; \tau_1)$ vanish, Eq.~\eqref{eqn:dual_killing_constraints} holds.
Of the 50 remaining components, only 31 are nonzero, which are given in Eq.~\eqref{eqn:dual_killing_comps}.
Note, however, that these 31 components are only determined by 12 functions, the independent components of the transverse Jacobi propagators.

\subsection{Observables from a spinning test particle}

The last observable discussed in~\cite{Flanagan2019} was an observable that an observer can measure using a spinning test particle.
The definition is as follows: consider an observer who moves along a geodesic $\gamma$, and a spinning test particle which moves along a curve $\bar{\gamma}$.
The two are initially comoving at some proper time $\tau_0$.
The observer measures her initial separation from the particle, the particle's initial momentum, and its initial intrinsic spin per unit mass.
The observer also performs these measurements at some later proper time $\tau_1$, and then compares the results with her initial measurements by parallel transporting the initial measurements to this final time.
The persistent observables are the differences between the final and initial measurements, which we denote by $\Delta \xi^{a'}_{\textrm{S}}$ for separation, $\Delta p^{a'}$ for momentum, and $\Delta s^{a'}$ for intrinsic spin per unit mass.

To describe the time evolution of a spinning test particle, one must ascribe to the particle a worldline that represents the center of mass of the particle (which is fixed by a \emph{spin supplementary condition}).
We use the Tulczyjew condition~\cite{Tulczyjew1959}, which is

\begin{equation}
  p_a j^{ab} = 0,
\end{equation}
where $p^a$ and $j^{ab}$ are the linear and angular momenta of the spinning particle, respectively.
The definition of intrinsic spin per unit mass is also fixed by this condition and is given by

\begin{equation}
  s^a \equiv -\frac{1}{2 p_e p^e} \epsilon^{abcd} p_b j_{cd}.
\end{equation}

Although the observables $\Delta \xi_{\textrm S}^{a'}$, $\Delta p^{a'}$, and $\Delta s^{a'}$ are, in general, nonlinear functions of initial separation, momentum, and spin, we will expand in the initial separation and spin as we did in our previous paper\footnote{Note that we wrote Eq.~\eqref{eqn:Delta_s_decomp} in~\cite{Flanagan2019} in terms of a new bitensor $\Sigma^{a'}{}_{bc}$.
  Here we avoid introducing $\Sigma^{a'}{}_{bc}$ by using the fact that it can be expressed in terms of the holonomy at low enough order in $s^a$.}

\begin{subequations} \label{eqn:spin_obs_decomp}
  \begin{align}
    \Delta \xi^{a'}_{\textrm S} &\equiv \Big[\Delta K^{a'}{}_b + L^{a'}{}_{bc} \xi^c + O(\boldsymbol{\xi}^2)\Big] \xi^b \nonumber \\
    &\hspace{2em}+ \left[\Upsilon^{a'}{}_b + \Psi^{a'}{}_{bc} \xi^c + O(\boldsymbol{\xi}^2)\right] s^b + O(\boldsymbol{s})^2, \displaybreak[0] \label{eqn:Delta_xi_decomp} \\
    \Delta p^{a'} &= \sqrt{-p_b p^b} \frac{\uD}{\ud \tau_1} \Delta \xi^{a'}_{\textrm S} + O(\boldsymbol{s})^2 \displaybreak[0] \label{eqn:Delta_p_decomp} \\
    \Delta s^{a'} &= -\pb{\gamma} g^{a'}{}_a \left[\Omega^a{}_b (\gamma, \bar{\gamma}; \tau_1) + O(\boldsymbol{\xi}, \dot{\boldsymbol \xi})^2\right] s^b + O(\boldsymbol{s})^2. \label{eqn:Delta_s_decomp}
  \end{align}
\end{subequations}

Expressions for these bitensors in general spacetimes were given in Eq.~(4.47) of~\cite{Flanagan2019} in terms of Jacobi and parallel propagators; they can be computed in plane wave spacetimes using that result and Eq.~\eqref{eqn:4d_propagators}.
Assuming that $\xi^a \ell_a = 0$, we find that

\begin{subequations} \label{eqn:spinning_results}
  \begin{align}
    \Upsilon^i{}_j (\tau_1, \tau_0) &= -\chi \pb{\dot x} [\Upsilon^i{}_u]_j (\tau_1) \nonumber \\
    &= \int_{u_0}^{u_1} (u_1 - u_2) \mat{H}^i{}_k (u_1, u_2) \left(\mat{\mathcal A}^*\right){}^k{}_j (u_2), \\
    \pb{x} [\Upsilon^{v'}{}_i]_j (\tau_0) &= -\chi \pb{x \dot x} [\Upsilon^{v'}{}_u]_{ij} \nonumber \\
    &= \partial_{u_1} \mat{K}_{kj} (u_1, u_0) \Upsilon^k{}_i (\tau_1, \tau_0), \displaybreak[0] \\
    \pb{\dot x} [\Upsilon^{v'}{}_i]_j (\tau_0) &= \frac{1}{\chi} \partial_{u_1} \left[(u_1 - u_0) \mat{H}_{kj} (u_1, u_0)\right] \Upsilon^k{}_i (\tau_1, \tau_0), \\
    \pb{\dot{x}^2} [\Upsilon^{v'}{}_u]_{ij} &= -\frac{1}{\chi} \pb{\dot x} [\Upsilon^{v'}{}_{(j}]_{i)} (\tau_0), \displaybreak[0] \\
    \Psi^{v'}{}_{ij} (\tau_0) &= -\chi \pb{\dot x} [\Psi^{v'}{}_{ui}]_j (\tau_0) \nonumber \\
    &= \int_{u_0}^{u_1} \ud u_2 (u_1 - u_2) \left(\mat{\mathcal A}^*\right){}_{ki} (u_2) \nonumber \\
    &\hspace{4em}\times \partial_{u_2} \mat{K}^k{}_j (u_2, u_0), \label{eqn:spin_obs_Psi}
  \end{align}
\end{subequations}
where

\begin{equation}
  (\mat{\mathcal A}^*)_{ij} (u) \equiv \mat{\mathcal A}_{ik} \mat{\epsilon}^k{}_j.
\end{equation}
We do not determine the spinning test particle observables nonperturbatively, since we are not aware of ways to solve the fully nonlinear Mathisson-Papapetrou equations in plane wave spacetimes.

At this point, let us focus on the observable $\Psi^{v'}{}_{ij} (\tau_0)$, which is an observable which does not seem able to be expressed solely in terms of sums and products of transverse Jacobi propagators and their derivatives.
Using Eqs.~\eqref{eqn:switch_identities} and~\eqref{eqn:spin_obs_Psi}, we can show that

\begin{equation} \label{eq:Psi_v_ij}
  \begin{split}
    \Psi^{v'}{}_{ij} &(\tau_0) \\
    &= -\partial_{u_0} \int_{u_0}^{u_1} \ud u_2 (u_1 - u_2) \mat{K}_j{}^k (u_0, u_2) \left(\mat{\mathcal A}^*\right){}_{ki} (u_2).
  \end{split}
\end{equation}
The integrand does not appear to be in the form of a total derivative [unless $\mat{K}_j{}^k (u_0, u_1)$ and $\mat{K}_j{}^k (u_1, u_0)$ are proportional by a constant, which is not necessarily true].
As in Sec.~\ref{sec:curve_dev}, we conclude by computing the proper time delay observable (but now for the spinning test particle):

\begin{equation}
  \begin{split}
    \dot{\gamma}_{a'} \Delta \xi_S^{a'} = -\chi\Big[&L^{v'}{}_{ij} (\tau_0) \xi^i (\tau_0) \xi^j (\tau_0) \\
    &+ \Psi^{v'}{}_{ij} (\tau_0) \xi^i (\tau_0) s^i (\tau_0) + O(\boldsymbol{\xi}, \boldsymbol{s}^2)\Big].
  \end{split}
\end{equation}
Thus, $\Psi^{v'}{}_{ij} (\tau_0)$, like $L^{v'}{}_{ij} (\tau_0)$ in Sec.~\ref{sec:curve_dev}, measures a sort of proper time delay observable, except that it gives the dependence of this delay on spin in addition to separation.

\subsection{Observables at second order in curvature}

As in Sec.~\ref{sec:2nd_jacobi}, we now compute some parts of our persistent observables at second order in curvature.
We do this both for general plane wave spacetimes and for the specific plane wave spacetime which we introduced in Sec.~\ref{sec:model_waveform}.
We focus on the quantities $L^{v'}{}_{ij} (\tau_0)$ in Eq.~\eqref{eqn:curve_dev_decomp} and $\Psi^{v'}{}_{ij} (\tau_0)$ in Eq.~\eqref{eq:Psi_v_ij} in this section.
These results illustrate features of observables which can be computed from the transverse Jacobi propagators and their derivatives; other such observables are qualitatively similar.

The first observable which we compute is $L^{v'}{}_{ij} (\tau_0)$, which is a piece of the curve deviation observable defined by Eq.~\eqref{eqn:curve_dev_decomp}.
The value of this observable in arbitrary plane wave spacetimes is given by Eq.~\eqref{eqn:curve_dev_results_L}.
Expanding this expression order-by-order, we find that it vanishes at zeroth order, whereas at first order we find it is

\begin{equation}
  \pp{(1)} L^{v'}{}_{ij} (\tau_0) = \frac{1}{2} \partial_{u_1} \pp{(1)} \mat{K}{}_{(ij)} (u_1, u_0),
\end{equation}
and at second order it is

\begin{widetext}
\begin{equation}
  \begin{split}
    \pp{(2)} L^{v'}{}_{ij} (\tau_0) = \frac{1}{2} \Bigg\{&\pp{(1)} \mat{K}{}_{k(i} (u_1, u_0) \partial_{u_1} \pp{(1)} \mat{K}^k{}_{j)} (u_1, u_0) + \frac{1}{2} \partial_{u_1} \left[\pp{(1)} \mat{K}{}_{ik} (u_1, u_0) \pp{(1)} \mat{K}^k{}_j (u_1, u_0)\right] \\
    &- \int_{u_0}^{u_1} \ud u_2 \int_{u_0}^{u_2} \ud u_3 \partial_{u_3} \pp{(1)} \mat{K}{}_{ik} (u_3, u_0) \pp{(1)} \mat{K}^k{}_j (u_3, u_0)\Bigg\}.
  \end{split}
\end{equation}
\end{widetext}
At second order, this observable is pure trace because it is symmetric and constructed from products of $\pp{(1)} \mat{K}^i{}_j (u', u)$, which is itself a symmetric and trace-free $2 \times 2$ matrix (assuming a vacuum plane wave spacetime).

Using the wave profile~\eqref{eqn:model_waveform}, we have that

\begin{equation}
  \begin{split}
    \left.L^{v'}{}_{ij} (0)\right|_{\tau_1 = \frac{2\pi n}{\omega \chi}} = &-\frac{\pi \omega n \epsilon^2}{2} \left\{\left[\cos (2\phi) - 1\right] a^2 + 3\right\} \delta_{ij} \\
    &+ O(\epsilon^3).
  \end{split}
\end{equation}
Note that, like $\partial_{u_1} A^i{}_j (u_1, u_0)$, this observable vanishes at first order in $\epsilon$.

The next observable which we consider is $\Psi^{v'}{}_{ij} (\tau_0)$, which is an observable from a spinning test particle which is defined by Eq.~\eqref{eqn:Delta_xi_decomp}.
This observable is vanishing at first order by Eq.~\eqref{eqn:spin_obs_Psi}, and this equation also implies that (at second order)

\begin{widetext}
\begin{equation}
  \pp{(2)} \Psi^{v'}{}_{ij} (\tau_0) = \int_{u_0}^{u_1} \ud u_2 \left\{\partial_{u_2} \pp{(1)} \mat{K}{}_{kl} (u_2, u_0) \partial_{u_2} \pp{(1)} \mat{K}^k{}_j (u_2, u_0) + \frac{1}{2} (u_1 - u_2) [\mat{\mathcal A} (u_2), \partial_{u_2} \mat{K} (u_2, u_0)]_{lj}\right\} \underline{\epsilon}^l{}_i. \\
\end{equation}
\end{widetext}
As with the transverse Jacobi propagators, at second order there are both pieces that are pure trace and pieces that are antisymmetric (assuming a vacuum plane wave).
However, because of the factor of $\underline{\epsilon}{}_{ab}$, it is the pure trace piece which only occurs when the wave is not linearly polarized, instead of the antisymmetric piece.

Finally, we consider the wave profile in Eq.~\eqref{eqn:model_waveform}; we find that

\begin{equation}
  \begin{split}
    \left.\Psi^{v'}{}_{ij} (0)\right|_{\tau_1 = \frac{2\pi n}{\omega \chi}} = &-2 \pi^2 \omega n^2 \epsilon^2 a \sqrt{1 - a^2} \sin \phi \delta_{ij} \\
    &+ \frac{\pi \omega n \epsilon^2}{2} \left\{\left[\cos(2\phi) - 1\right] a^2 + 3\right\} \underline{\epsilon}{}_{ij} \\
    &+ O(\epsilon^3).
  \end{split}
\end{equation}
This expression has the same qualitative features as in the case of a general wave profile.

\section{Discussion} \label{sec:discussion}

In this paper, we have investigated the behavior of the persistent gravitational wave observables of~\cite{Flanagan2019} in nonlinear, exact plane wave spacetimes.
These spacetimes possess an important set of two functions (and their derivatives), which we refer to as transverse Jacobi propagators.
Many of the geometric properties of these spacetimes, such as Killing vectors and solutions to the geodesic equation, can be written in terms of these functions.
Our primary result is that many parts of the observables introduced in~\cite{Flanagan2019} can be determined just from the values of these functions and their derivatives.
We found in our linear, plane wave results in~\cite{Flanagan2019} that many parts of our observables could be written in terms of a small number of functions, but the fact that this statement also holds in the nonlinear context is unexpected.

The main utility of this result is that only the transverse Jacobi propagators are necessary to determine the values of many of our persistent observables.
That is, although the persistent observables we have defined in~\cite{Flanagan2019} encompass a large number of interesting physical effects, many of these effects are determined by just a small number of functions.
These functions, in turn, can be determined by the displacement memory observable (which gives the transverse Jacobi propagators directly) and the relative velocity observable (which gives their derivatives).

Finally, we conclude with a few remarks about extending our results to the class of ``\emph{pp}-wave'' spacetimes, which are a generalization of plane wave spacetimes where the planar wave fronts are not homogeneous, as $\mat{\mathcal A}{}_{ij}$ is also a function of $x^i$.
Such spacetimes have Jacobi and parallel propagators that one can calculate using a procedure that is similar to the one we carried out in this paper, but the geodesic equation does not have exact solutions, nor are the transverse Jacobi propagators solely functions of $u$.
In plane wave spacetimes, one only needed to determine the transverse Jacobi propagators along a given timelike geodesic in order to compute persistent observables, but in \emph{pp}-wave spacetimes one would need to determine them along all timelike geodesics.

\section*{Acknowledgments}

\'E. \'E. F. and A. M. G. acknowledge the support of NSF Grant No. PHY-1707800 to Cornell University.
D.A.N. acknowledges the support of the Netherlands Organization for Scientific Research through the NWO VIDI Grant No. 639.042.612-Nissanke.

\appendix

\section{Values of the Holonomy Observables} \label{app:holonomy}

In this appendix, we give the values that our holonomy observables take in plane wave spacetimes.

We start with the holonomy observable $\zero{\Omega}{}^A{}_B (\gamma, \bar{\gamma}; \tau')$ for affine transport [$\varkappa = (0, 0, 0, 0)$].
First, the components of $\smallunderset{PP}{\zero \Omega}{}^a{}_b (\gamma, \bar{\gamma}; \tau')$ are determined from Eq.~\eqref{eqn:0_holonomy_pp} to be
\begin{equation}
\smallunderset{PP}{\zero \Omega}{}^a{}_b (\gamma, \bar{\gamma}; \tau_1) = \Omega^a{}_b (\gamma, \bar{\gamma}; \tau_1).
\end{equation}
From Eqs.~\eqref{eqn:0_holonomy_assembly}, \eqref{eqn:holonomy_xi}, and~\eqref{eqn:chi_xi}, it is possible to show that the remaining components of $\zero{\Omega}{}^A{}_B (\gamma, \bar{\gamma}; \tau_1)$ are

\begin{subequations} \label{eqn:affine_comps}
  \begin{align}
    \smallunderset{JP}{\zero \Omega}{}^{uv}{}_u (\gamma, \bar{\gamma}; \tau_1) &= -\Delta \chi^v (\gamma, \bar{\gamma}; \tau_1), \\
    \smallunderset{JP}{\zero \Omega}{}^{ui}{}_u (\gamma, \bar{\gamma}; \tau_1) &= \smallunderset{JP}{\zero \Omega}{}^{vi}{}_v (\gamma, \bar{\gamma}; \tau_1) = -\Delta \chi^i (\gamma, \bar{\gamma}; \tau_1, \tau_0), \displaybreak[0] \\
    \smallunderset{JP}{\zero \Omega}{}^{vi}{}_u (\gamma, \bar{\gamma}; \tau_1) &= \Delta \chi^v (\gamma, \bar{\gamma}; \tau_1) \Omega^i{}_u (\gamma, \bar{\gamma}; \tau_1) \nonumber \\
    &\hspace{1em}- \Delta \chi^i (\gamma, \bar{\gamma}; \tau_1, \tau_0) \Omega^v{}_u (\gamma, \bar{\gamma}; \tau_1), \displaybreak[0] \\
    \smallunderset{JP}{\zero \Omega}{}^{vi}{}_j (\gamma, \bar{\gamma}; \tau_1) &= \Delta \chi^v (\gamma, \bar{\gamma}; \tau_1) \delta^i{}_j \nonumber \\
    &\hspace{1em}- \Delta \chi^i (\gamma, \bar{\gamma}; \tau_1, \tau_0) \Omega^v{}_j (\gamma, \bar{\gamma}; \tau_1), \displaybreak[0] \\
    \smallunderset{JP}{\zero \Omega}{}^{ij}{}_u (\gamma, \bar{\gamma}; \tau_1) &= 2 \Delta \chi^{[i} (\gamma, \bar{\gamma}; \tau_1, \tau_0) \Omega^{j]}{}_u (\gamma, \bar{\gamma}; \tau_1), \displaybreak[0] \\
    \smallunderset{JP}{\zero \Omega}{}^{ij}{}_k (\gamma, \bar{\gamma}; \tau_1, \tau_0) &= 2 \Delta \chi^{[i} (\gamma, \bar{\gamma}; \tau_1, \tau_0) \delta^{j]}{}_k, \displaybreak[0] \\
    \smallunderset{JJ}{\zero \Omega}{}^{uv}{}_{ui} (\gamma, \bar{\gamma}; \tau_1) &= \frac{1}{2} \Omega^v{}_i (\gamma, \bar{\gamma}; \tau_1), \\
    \smallunderset{JJ}{\zero \Omega}{}^{vi}{}_{uv} (\gamma, \bar{\gamma}; \tau_1) &= -\frac{1}{2} \Omega^i{}_u (\gamma, \bar{\gamma}; \tau_1), \\
    \smallunderset{JJ}{\zero \Omega}{}^{vi}{}_{uj} (\gamma, \bar{\gamma}; \tau_1) &= \frac{1}{2} \Big[\delta^i{}_j \Omega^v{}_u (\gamma, \bar{\gamma}; \tau_1) \nonumber \\
    &\hspace{2.5em}- \Omega^i{}_u (\gamma, \bar{\gamma}; \tau_1) \Omega^v{}_i (\gamma, \bar{\gamma}; \tau_1)\Big], \\
    \smallunderset{JJ}{\zero \Omega}{}^{vi}{}_{jk} (\gamma, \bar{\gamma}; \tau_1) &= -\delta^i{}_{[j} \Omega^v{}_{k]} (\gamma, \bar{\gamma}; \tau_1), \\
    \smallunderset{JJ}{\zero \Omega}{}^{ij}{}_{uk} (\gamma, \bar{\gamma}; \tau_1) &= -\delta^{[i}{}_k \Omega^{j]}{}_u (\gamma, \bar{\gamma}; \tau_1).
  \end{align}
\end{subequations}
These expressions still involve the components of $\Omega^a{}_b (\gamma, \bar{\gamma}; \tau_1)$ and $\Delta \chi^a (\gamma, \bar{\gamma}; \tau_1)$.
We now write each of these quantities as an expansion in $\xi^a$ and $\dot{\xi}^a$:

\begin{widetext}
\begin{subequations} \label{eqn:affine_expansion}
  \begin{align}
    \Omega^a{}_b (\gamma, \bar{\gamma}; \tau_1) &\equiv \sum_{m = 1}^\infty \sum_{k = 0}^m \pb{\xi^k \dot{\xi}^{m - k}} \Omega^a{}_{bc_1 \cdots c_k d_1 \cdots d_{m - k}} (\tau_1) \xi^{c_1} \cdots \xi^{c_k} \dot{\xi}^{d_1} \cdots \dot{\xi}^{d_{m - k}}, \\
    \Delta \chi^a (\gamma, \bar{\gamma}; \tau_1) &\equiv \sum_{m = 1}^\infty \sum_{k = 0}^m \pb{\xi^k \dot{\xi}^{m - k}} \Delta \eta^a{}_{b_1 \cdots b_k c_1 \cdots c_{m - k}} (\tau_1) \xi^{b_1} \cdots \xi^{b_k} \dot{\xi}^{c_1} \cdots \dot{\xi}^{c_{m - k}}.
  \end{align}
\end{subequations}
The nonzero components of the coefficients in this expansion are given by Eqs.~\eqref{eqn:holonomy_xi} and~\eqref{eqn:chi_xi}:

\begin{subequations} \label{eqn:affine_coeffs}
  \begin{align}
    \pb{\xi} \Omega^v{}_{ij} (\tau_1) &= \pb{\xi} \Omega_{iuj} (\tau_1) = -\partial_{u_1} \mat{K}{}_{ij} (u_1, u_0), \\
    \pb{\dot \xi} \Omega^v{}_{ij} (\tau_1) &= \pb{\dot \xi} \Omega_{iuj} (\tau_1) = -\frac{1}{\chi} \left\{\partial_{u_1} \left[(u_1 - u_0) \mat{H}{}_{ij} (u_1, u_0)\right] - \delta_{ij}\right\}, \\
    \pb{\xi^2} \Omega^v{}_{uij} (\tau_1) &= \frac{1}{2} \partial_{u_1} \mat{K}{}_{k(i} (u_1, u_0) \partial_{u_1} \mat{K}^k{}_{j)} (u_1, u_0), \\
    \pb{\xi \dot \xi} \Omega^v{}_{uij} (\tau_1) &= \frac{1}{\chi} \partial_{u_1} \mat{K}{}_{ki} (u_1, u_0) \left\{\partial_{u_1} \left[(u_1 - u_0) \mat{H}^k{}_j (u_1, u_0)\right] - \delta^k{}_j\right\}, \\
    \pb{\dot{\xi}^2} \Omega^v{}_{uij} (\tau_1) &= \frac{1}{2\chi^2} \left\{\partial_{u_1} \left[(u_1 - u_0) \mat{H}{}_{k(i} (u_1, u_0)\right] - \delta_{k(i}\right\} \left\{\partial_{u_1} \left[(u_1 - u_0) \mat{H}^k{}_{j)} (u_1, u_0)\right] - \delta^k{}_{j)}\right\}, \displaybreak[0] \\
    \pb{\xi} \Delta \eta^i{}_j (\tau_1, \tau_0) &= \delta^i{}_j - \left[\mat{K}^i{}_j (u_1, u_0) - (u_1 - u_0) \partial_{u_1} \mat{K}^i{}_j (u_1, u_0)\right], \\
    \pb{\dot \xi} \Delta \eta^i{}_j (\tau_1, \tau_0) &= \frac{1}{\chi} (u' - u)^2 \partial_{u_1} \mat{H}^i{}_j (u_1, u_0), \\
    \pb{\dot x} [\pb{\xi} \Delta \eta^v{}_i]_j (\tau_1) &= \frac{1}{\chi} \pb{\xi} \Delta \eta_{ji} (\tau_1, \tau_0), \\
    \pb{\dot x} [\pb{\dot \xi} \Delta \eta^v{}_i]_j (\tau_1) &= \frac{1}{\chi} \pb{\dot \xi} \Delta \eta_{ji} (\tau_1, \tau_0), \\
    \pb{\xi^2} \Delta \eta^v{}_{ij} (\tau_1) &= \frac{1}{2} \partial_{u_1} A_{k(i} (u_1, u_0) \left[\mat{K}^k{}_{j)} (u_1, u_0) - (u_1 - u_0) \partial_{u_1} \mat{K}^k{}_{j)} (u_1, u_0)\right], \\
    \pb{\xi \dot{\xi}} \Delta \eta^v{}_{ij} (\tau_1) &= \frac{1}{2 \chi} \Big\{\left[\mat{K}^k{}_i (u_1, u_0) - (u_1 - u_0) \partial_{u_1} \mat{K}^k{}_i (u_1, u_0)\right] \left(\partial_{u_1} \left[(u_1 - u_0) \mat{H}{}_{kj} (u_1, u_0)\right] - \delta_{kj}\right) \nonumber \\
    &\hspace{3.4em}- (u_1 - u_0)^2 \partial_{u_1} \mat{K}{}_{ki} (u_1, u_0) \partial_{u_1} \mat{H}^k{}_j (u_1, u_0) + \delta_{ij} - \left[\mat{K}_{ji} (u_1, u_0) - (u_1 - u_0) \partial_{u_1} \mat{K}_{ji} (u_1, u_0)\right]\Big\}, \\
    \pb{\dot{\xi}^2} \Delta \eta^v{}_{ij} (\tau_1) &= -\frac{1}{\chi^2} (u_1 - u_0)^2 \left\{\frac{1}{2} \partial_{u_1} \left[(u_1 - u_0) \mat{H}{}_{k(i} (u_1, u_0)\right] - \delta_{k(i}\right\} \partial_{u_1} \mat{H}^k{}_{j)} (u_1, u_0).
  \end{align}
\end{subequations}

We now perform a similar calculation for the dual Killing holonomy, which is given in terms of the tensors that occur in Eq.~\eqref{eqn:dual_killing_affine_core}.
The components of $\Delta^A{}_B (\gamma; \tau_1)$ and $\hat{\Delta}^A{}_B (\gamma, \gamma; \tau_1)$ are given by a lengthy calculation starting with Eqs.~\eqref{eqn:4d_propagators}.
Their components are given by

\begin{subequations}
  \begin{align}
    \smallunderset{PP}{\Delta}^i{}_j (\gamma; \tau_1, \tau_0) &= \chi \pb{\dot x} [\smallunderset{PP}{\Delta}^v{}_j]^i (\gamma; \tau_1) = -\chi \pb{\dot x} [\smallunderset{PP}{\Delta}^i{}_u]_j (\gamma; \tau_1) = \mat{K}{}_j{}^i (u_0, u_1) - \delta_j{}^i, \\
    \pb{\dot{x}^2} [\smallunderset{PP}{\Delta}^v{}_u]_{ij} (\gamma; \tau_1) &= -\frac{1}{\chi^2} \smallunderset{PP}{\Delta}{}_{(ij)} (\gamma; \tau_1, \tau_0), \displaybreak[0] \\
    \smallunderset{PJ}{\Delta}^i{}_{uj} (\gamma; \tau_1) &= \chi \pb{\dot x} [\smallunderset{PJ}{\Delta}^v{}_{uj}]^i (\gamma; \tau_1) \nonumber \\
    &= -\frac{1}{2} \left\{\mat{K}{}_k{}^i (u_0, u_1) \partial_{u_1} \mat{K}{}_j{}^k (u_1, u_0) + \partial_{u_0} \mat{K}{}_k{}^i \partial_{u_1} \left[(u_1 - u_0) (\mat{H}{}_j{}^k (u_1, u_0) - \delta_j{}^k)\right]\right\}, \displaybreak[0] \\
    \smallunderset{JP}{\Delta}^{ui}{}_j (\gamma; \tau_1) &= \chi \pb{\dot x} [\smallunderset{JP}{\Delta}^{uv}{}_j]^i (\gamma; \tau_1) = -\chi \pb{\dot x} [\smallunderset{JP}{\Delta}^{ui}{}_u]_j (\gamma; \tau_1) = 2 \chi^2 [\smallunderset{JP}{\Delta}^{vi}{}_j] (\gamma; \tau_1) = -2 \chi^3 \pb{\dot x} [\smallunderset{JP}{\Delta}^{vi}{}_u]_j (\gamma; \tau_1) \nonumber \\
    &= (u_1 - u_0) \left[\mat{K}{}_j{}^i (u_0, u_1) - \mat{H}{}_j{}^i (u_0, u_1)\right], \\
    \pb{\dot{x}^2} [\smallunderset{JP}{\Delta}^{uv}{}_u]_{ij} (\gamma; \tau_1) &= -\frac{1}{\chi^2} \smallunderset{JP}{\Delta}^u{}_{(ij)} (\gamma; \tau_1), \\
    \pb{\dot x} [\smallunderset{JP}{\Delta}^{ij}{}_k]_l (\gamma; \tau_1) &= \frac{2}{\chi} \delta^{[i}{}_l \smallunderset{JP}{\Delta}^{|u|j]}{}_k (\gamma; \tau_1), \\
    \pb{\dot{x}^2} [\smallunderset{JP}{\Delta}^{ij}{}_u]_{kl} (\gamma; \tau_1) &= -\frac{1}{\chi} \pb{\dot x} [\smallunderset{JP}{\Delta}^{ij}{}_{(k}]_{l)} (\gamma; \tau_1), \\
    \pb{\dot{x}^2} [\smallunderset{JP}{\Delta}^{vi}{}_j]_{kl} (\gamma; \tau_1) &= \delta_{kl} [\smallunderset{JP}{\Delta}^{vi}{}_j] (\gamma; \tau_1) - 2 \delta^i{}_{(k} [\smallunderset{JP}{\Delta}^v{}_{l)j}] (\gamma; \tau_1), \\
    \pb{\dot{x}^3} [\smallunderset{JP}{\Delta}^{vi}{}_u]_{jkl} (\gamma; \tau_1) &= -\frac{1}{\chi} \pb{\dot{x}^2} [\smallunderset{JP}{\Delta}^{vi}{}_{(j}]_{kl)} (\gamma; \tau_1), \displaybreak[0] \\
    \smallunderset{JJ}{\Delta}^{ui}{}_{uj} (\gamma; \tau_1) &= \chi \pb{\dot x} [\smallunderset{JJ}{\Delta}^{uv}{}_{uj}]^i (\gamma; \tau_1) = 2 \chi^2 [\smallunderset{JJ}{\Delta}^{vi}{}_{uj}] (\gamma; \tau_1) \nonumber \\
    &= \frac{1}{2} (u_1 - u_0) \left[\mat{H}{}_k{}^i (u_0, u_1) - \mat{K}{}_k{}^i (u_0, u_1)\right] \partial_{u_1} \mat{K}{}_j{}^k (u_1, u_0) \nonumber \\
    &\hspace{1em}+ \frac{1}{2} \left[(u_1 - u_0) \partial_{u_1} \mat{K}^i{}_k (u_1, u_0) - \mat{K}^i{}_k (u_1, u_0)\right] \partial_{u_1} \left\{(u_1 - u_0) \left[\mat{H}{}_j{}^k (u_1, u_0) - \delta_j{}^k\right]\right\}, \\
    \pb{\dot \xi} [\smallunderset{JJ}{\Delta}^{ij}{}_{uk}]_l (\gamma; \tau_1) &= \frac{2}{\chi} \delta^{[i}{}_l \smallunderset{JJ}{\Delta}^{|u|j]}{}_{uk} (\gamma; \tau_1), \\
    \pb{\dot{x}^2} [\smallunderset{JJ}{\Delta}^{vi}{}_{uj}]_{kl} (\gamma; \tau_1) &= \delta_{kl} [\smallunderset{JJ}{\Delta}^{vi}{}_{uj}] (\gamma; \tau_1) - 2 \delta^i{}_{(k} [\smallunderset{JJ}{\Delta}^v{}_{l)uj}] (\gamma; \tau_1),
  \end{align}
\end{subequations}
and

\begin{subequations}
  \begin{align}
    \smallunderset{PP}{\hat \Delta}{}^i{}_j (\gamma, \gamma; \tau_1, \tau_0) &= \chi \pb{\dot x} [\smallunderset{PP}{\hat \Delta}{}^v{}_j]^i (\gamma, \gamma; \tau_1) = -\chi \pb{\dot x} [\smallunderset{PP}{\hat \Delta}{}^i{}_u]_j (\gamma, \gamma; \tau_1) \nonumber \\
    &= \mat{K}{}_j{}^i (u_1, u_0) - \delta_j{}^i - (u_1 - u_0) \partial_{u_1} \mat{K}{}_j{}^i (u_1, u_0), \\
    \pb{\dot{x}^2} [\smallunderset{PP}{\hat \Delta}{}^v{}_u]_{ij} (\gamma, \gamma; \tau_1) &= -\frac{1}{\chi^2} \smallunderset{PP}{\hat \Delta}{}_{(ij)} (\gamma, \gamma; \tau_1, \tau_0), \displaybreak[0] \\
    \smallunderset{PJ}{\hat \Delta}{}^i{}_{uj} (\gamma, \gamma; \tau_1) &= \chi \pb{\dot x} [\smallunderset{PJ}{\hat \Delta}{}^v{}_{uj}]^i (\gamma, \gamma; \tau_1) = \frac{1}{2} \partial_{u_1} \mat{K}{}_j{}^i (u_1, u_0), \displaybreak[0] \\
    \smallunderset{JP}{\hat \Delta}{}^{ui}{}_j (\gamma, \gamma; \tau_1) &= \chi \pb{\dot x} [\smallunderset{JP}{\hat \Delta}{}^{uv}{}_j]^i (\gamma, \gamma; \tau_1) = -\chi \pb{\dot x} [\smallunderset{JP}{\hat \Delta}{}^{ui}{}_u]_j (\gamma, \gamma; \tau_1) = 2 \chi^2 [\smallunderset{JP}{\hat \Delta}{}^{vi}{}_j] (\gamma, \gamma; \tau_1) = -2 \chi^3 [\smallunderset{JP}{\hat \Delta}{}^{vi}{}_u]_j (\gamma, \gamma; \tau_1) \nonumber \\
    &= -(u_1 - u_0)^2 \partial_{u_1} \mat{H}{}_j{}^i (u_1, u_0), \\
    \pb{\dot{x}^2} [\smallunderset{JP}{\hat \Delta}{}^{uv}{}_u]_{ij} (\gamma, \gamma; \tau_1) &= -\frac{1}{\chi^2} \smallunderset{JP}{\hat \Delta}{}^u{}_{(ij)} (\gamma, \gamma; \tau_1), \\
    \pb{\dot x} [\smallunderset{JP}{\hat \Delta}{}^{ij}{}_k]_l (\gamma, \gamma; \tau_1, \tau_0) &= \frac{2}{\chi} \delta^{[i}{}_l \smallunderset{JP}{\hat \Delta}{}^{|u|j]}{}_k (\gamma, \gamma; \tau_1), \\
    \pb{\dot{x}^2} [\smallunderset{JP}{\hat \Delta}{}^{ij}{}_u]_{kl} (\gamma, \gamma; \tau_1) &= -\frac{1}{\chi} \pb{\dot x} [\smallunderset{JP}{\hat \Delta}{}^{ij}{}_{(k}]_{l)} (\gamma, \gamma; \tau_1, \tau_0), \\
    \pb{\dot{x}^2} [\smallunderset{JP}{\hat \Delta}{}^{vi}{}_j]_{kl} (\gamma, \gamma; \tau_1) &= \delta_{kl} [\smallunderset{JP}{\hat \Delta}{}^{vi}{}_j] (\gamma, \gamma; \tau_1) - 2 \delta^i{}_{(k} [\smallunderset{JP}{\hat \Delta}{}^v{}_{l)j}] (\gamma, \gamma; \tau_1), \\
    \pb{\dot{x}^3} [\smallunderset{JP}{\hat \Delta}{}^{vi}{}_u]_{jkl} (\gamma, \gamma; \tau_1) &= -\frac{1}{\chi} \pb{\dot{x}^2} [\smallunderset{JP}{\hat \Delta}{}^{vi}{}_{(j}]_{kl)} (\gamma, \gamma; \tau_1), \displaybreak[0] \\
    \smallunderset{JJ}{\hat \Delta}{}^{ui}{}_{uj} (\gamma, \gamma; \tau_1) &= \chi \pb{\dot x} [\smallunderset{JJ}{\hat \Delta}{}^{uv}{}_{uj}]^i (\gamma, \gamma; \tau_1) = 2 \chi^2 [\smallunderset{JJ}{\hat \Delta}{}^{vi}{}_{uj}] (\gamma, \gamma; \tau_1) = \frac{1}{2} \partial_{u_1} \left\{(u_1 - u_0) \left[\mat{H}{}_j{}^i (u_1, u_0) - \delta_j{}^i\right]\right\}, \\
    \pb{\dot x} [\smallunderset{JJ}{\hat \Delta}{}^{ij}{}_{uk}]_l (\gamma, \gamma; \tau_1) &= \frac{2}{\chi} \delta^{[i}{}_l \smallunderset{JJ}{\hat \Delta}{}^{|u|j]}{}_{uk} (\gamma, \gamma; \tau_1), \\
    \pb{\dot{x}^2} [\smallunderset{JJ}{\hat \Delta}{}^{vi}{}_{uj}]_{kl} (\gamma, \gamma; \tau_1) &= \delta_{kl} [\smallunderset{JJ}{\hat \Delta}{}^{vi}{}_{uj}] (\gamma, \gamma; \tau_1) - 2 \delta^i{}_{(k} [\smallunderset{JJ}{\hat \Delta}{}^v{}_{l)uj}] (\gamma, \gamma; \tau_1),
  \end{align}
\end{subequations}
\end{widetext}
respectively.

There are now two additional pieces of notation that we must introduce before determining the nonzero components of Eq.~\eqref{eqn:dual_killing_affine_core}: first, we expand $\hat{\Delta}^A{}_B (\gamma, \bar{\gamma}; \tau_1)$ in powers of the separation as

\begin{equation}
  \hat{\Delta}^A{}_B (\gamma, \bar{\gamma}; \tau_1) \equiv \sum_{n = 0}^\infty \pb{\xi^n} \Theta^A{}_{Bc_1 \cdots c_n} (\gamma, \bar{\gamma}; \tau_1) \xi^{c_1} \cdots \xi^{c_n}.
\end{equation}
Next, it happens that in plane wave spacetimes, the components of the coefficients in this expansion depend on the sum $\dot{\bar x}^i (\tau_0) = \dot{x}^i (\tau_0) + \dot{\xi}^i (\tau_0)$, and not independently on either $\dot{x}^i (\tau_0)$ or $\dot{\xi}^i (\tau_0)$.
As such, we write [in analogy to Eq.~\eqref{eqn:polynomial_decomp} above] such quantities in terms of coefficients of the following expansion:

\begin{equation}
  Q^{\cdots}{}_{\cdots} = \sum_{k = 0}^n \pb{\dot{\bar x}^k} [Q^{\cdots}{}_{\cdots}]_{i_1 \cdots i_k} \dot{\bar x}^{i_1} (\tau_0) \cdots \dot{\bar x}^{i_k} (\tau_0).
\end{equation}
In this notation, we can show that the components of $\Theta^A{}_B (\gamma, \bar{\gamma}; \tau_1)$ and $\hat{\Delta}^A{}_B (\gamma, \gamma; \tau_1)$ are related (using $\Gamma$ and $\Delta$ for Brinkmann coordinate indices on the linear and angular momentum bundle):

\begin{equation}
  \pb{\dot{\bar x}^k} [\Theta^\Gamma{}_\Delta] (\gamma, \bar{\gamma}; \tau_1) = \pb{\dot{x}^k} [\hat{\Delta}^\Gamma{}_\Delta] (\gamma, \gamma; \tau_1).
\end{equation}

Using this notation, the final nonzero components that are needed are also given by a lengthy computation involving Eqs.~\eqref{eqn:4d_propagators}:

\begin{widetext}
\begin{subequations}
  \begin{align}
    \pb{\xi} \smallunderset{PP}{\Theta}^i{}_{uj} (\gamma, \bar{\gamma}; \tau_1) &= \chi \pb{\dot{\bar x}} [\pb{\xi} \smallunderset{PP}{\Theta}^v{}_{uj}]^i (\gamma, \bar{\gamma}; \tau_1) = \partial_{u'} \mat{K}{}_j{}^i (u_1, u_0), \displaybreak[0] \\
    \pb{\xi} \smallunderset{JP}{\Theta}^{vi}{}_{jv} (\gamma, \bar{\gamma}; \tau_1) &= -\chi \pb{\dot{\bar x}} [\pb{\xi} \smallunderset{JP}{\Theta}^{vi}{}_{uv}]_j (\gamma, \bar{\gamma}; \tau_1) = \mat{K}{}_j{}^i (u_1, u_0) - \delta_j{}^i - (u_1 - u_0) \partial_{u_1} \mat{K}{}_j{}^i (u_1, u_0), \\
    \pb{\dot{\bar x}} [\pb{\xi} \smallunderset{JP}{\Theta}^{iv}{}_{jk}]_l (\gamma, \bar{\gamma}; \tau_1) &= \frac{1}{\chi} \delta^i{}_k \pb{\xi} \smallunderset{JP}{\Theta}^v{}_{ljv} (\gamma, \bar{\gamma}; \tau_1), \\
    \pb{\xi} \smallunderset{JP}{\Theta}^{ij}{}_{kl} (\gamma, \bar{\gamma}; \tau_1, \tau_0) &= 2 \chi \pb{\dot{\bar x}} [\pb{\xi} \smallunderset{JP}{\Theta}^{[i|v|}{}_{kl}]^{j]} (\gamma, \bar{\gamma}; \tau_1), \\
    \pb{\xi} \smallunderset{JP}{\Theta}^{ui}{}_{uj} (\gamma, \bar{\gamma}; \tau_1) &= \chi \pb{\dot{\bar x}} [\pb{\xi} \smallunderset{JP}{\Theta}^{uv}{}_{uj}] (\gamma, \bar{\gamma}; \tau_1) = 2 \chi^2 [\pb{\xi} \smallunderset{JP}{\Theta}^{vi}{}_{uj}] (\gamma, \bar{\gamma}; \tau_1) = \partial_{u_1} \left\{(u_1 - u_0) \left[\mat{H}{}_j{}^i (u_1, u_0) - \delta_j{}^i\right]\right\}, \\
    \pb{\dot{\bar x}} [\pb{\xi} \smallunderset{JP}{\Theta}^{ij}{}_{uk}]_l (\gamma, \bar{\gamma}; \tau_1) &= -\frac{1}{\chi} \pb{\xi} \smallunderset{JP}{\Theta}^{ij}{}_{lk} (\gamma, \bar{\gamma}; \tau_1, \tau_0) + \frac{2}{\chi} \delta^{[i}{}_l \pb{\xi} \smallunderset{JP}{\Theta}^{|u|j]}{}_{uk} (\gamma, \bar{\gamma}; \tau_1), \\
    \pb{\dot{\bar x}^2} [\pb{\xi} \smallunderset{JP}{\Theta}^{vi}{}_{uj}]_{kl} (\gamma, \bar{\gamma}; \tau_1) &= -\frac{1}{\chi} \pb{\dot{\bar x}} [\pb{\xi} \smallunderset{JP}{\Theta}^{iv}{}_{(k|j|}]_{l)} (\gamma, \bar{\gamma}; \tau_1) + \delta_{kl} [\pb{\xi} \smallunderset{JP}{\Theta}^{vi}{}_{uj}] (\gamma, \bar{\gamma}; \tau_1) - 2 \delta^i{}_{(k} [\pb{\xi} \smallunderset{JP}{\Theta}^v{}_{l)uj}] (\gamma, \bar{\gamma}; \tau_1), \displaybreak[0] \\
    \pb{\xi} \smallunderset{JJ}{\Theta}^{vi}{}_{ujv} (\gamma, \bar{\gamma}; \tau_1) &= \frac{1}{2} \partial_{u_1} \mat{K}{}_j{}^i (u_1, u_0), \\
    \pb{\dot{\bar x}} [\pb{\xi} \smallunderset{JJ}{\Theta}^{iv}{}_{ujk}]_l (\gamma, \bar{\gamma}; \tau_1) &= \frac{1}{\chi} \delta^i{}_k \smallunderset{JJ}{\Theta}^v{}_{lujv} (\gamma, \bar{\gamma}; \tau_1), \\
    \pb{\xi} \smallunderset{JJ}{\Theta}^{ij}{}_{ukl} (\gamma, \bar{\gamma}; \tau_1) &= 2 \chi \pb{\dot{\bar x}} [\pb{\xi} \smallunderset{JJ}{\Theta}^{[i|v|}{}_{ukl}]^{j]} (\gamma, \bar{\gamma}; \tau_1).
  \end{align}
\end{subequations}
For our final result, we first define

\begin{equation}
  \Xi^A{}_B (\gamma, \bar{\gamma}; \tau_1) \equiv \zero{\Omega}{}^A{}_B (\gamma, \bar{\gamma}; \tau_1) + \hat{\Delta}{}^A{}_B (\gamma, \bar{\gamma}; \tau_1) - \hat{\Delta}{}^A{}_B (\gamma, \gamma; \tau_1);
\end{equation}
then Eq.~\eqref{eqn:dual_killing_affine_core} implies that

\begin{subequations} \label{eqn:dual_killing_comps}
  \begin{align}
    \smallunderset{PP}{\half \Omega}{}^v{}_u (\gamma, \bar{\gamma}; \tau_1) &= \smallunderset{PP}{\Xi}^v{}_u (\gamma, \bar{\gamma}; \tau_1) + \smallunderset{PP}{\Xi}^v{}_i (\gamma, \bar{\gamma}; \tau_1) \smallunderset{PP}{\Delta}{}^i{}_u (\gamma; \tau_1) + \smallunderset{PP}{\hat \Delta}{}^v{}_i (\gamma, \bar{\gamma}; \tau_1) \smallunderset{PP}{\zero \Omega}{}^i{}_u (\gamma, \bar{\gamma}; \tau_1) \nonumber \\
    &\hspace{1em}+ 2 \left[\smallunderset{PJ}{\hat \Delta}{}^v{}_{ui} (\gamma, \bar{\gamma}; \tau_1) - \smallunderset{PJ}{\hat \Delta}{}^v{}_{ui} (\gamma, \gamma; \tau_1)\right] \smallunderset{JP}{\Delta}^{ui}{}_u (\gamma; \tau_1) + 2 \smallunderset{PJ}{\hat \Delta}{}^v{}_{ui} (\gamma, \bar{\gamma}; \tau_1) \smallunderset{JP}{\zero \Omega}{}^{ui}{}_u (\gamma, \bar{\gamma}; \tau_1), \\
    \smallunderset{PP}{\half \Omega}{}^v{}_i (\gamma, \bar{\gamma}; \tau_1) &= \smallunderset{PP}{\Xi}^v{}_j (\gamma, \bar{\gamma}; \tau_1) \left[\delta^j{}_i + \smallunderset{PP}{\Delta}^j{}_i (\gamma; \tau_1)\right] + 2 \left[\smallunderset{PJ}{\hat \Delta}{}^v{}_{uj} (\gamma, \bar{\gamma}; \tau_1) - \smallunderset{PJ}{\hat \Delta}{}^v{}_{uj} (\gamma, \gamma; \tau_1)\right] \smallunderset{JP}{\Delta}^{uj}{}_i (\gamma; \tau_1), \displaybreak[0] \\
    \smallunderset{PJ}{\half \Omega}{}^v{}_{ui} (\gamma, \bar{\gamma}; \tau_1) &= \smallunderset{PP}{\Xi}{}^v{}_j (\gamma, \bar{\gamma}; \tau_1) \smallunderset{PP}{\Delta}^j{}_{ui} (\gamma; \tau_1) + 2 \left[\smallunderset{PJ}{\hat \Delta}{}^v{}_{uj} (\gamma, \bar{\gamma}; \tau_1) - \smallunderset{PJ}{\hat \Delta}{}^v{}_{uj} (\gamma, \gamma; \tau_1)\right] \smallunderset{JJ}{\Delta}^{uj}{}_{ui} (\gamma; \tau_1) \nonumber \\
    &\hspace{1em}+ \smallunderset{PJ}{\hat \Delta}{}^v{}_{ui} (\gamma, \bar{\gamma}; \tau_1) - \smallunderset{PJ}{\hat \Delta}{}^v{}_{ui} (\gamma, \gamma; \tau_1), \displaybreak[0] \\
    \smallunderset{JP}{\half \Omega}{}^{uv}{}_u (\gamma, \bar{\gamma}; \tau_1) &= \smallunderset{JP}{\Xi}^{uv}{}_u (\gamma, \bar{\gamma}; \tau_1) + 2 \smallunderset{JJ}{\Xi}^{uv}{}_{ui} (\gamma, \bar{\gamma}; \tau_1) \smallunderset{JP}{\Delta}^{ui}{}_u (\gamma; \tau_1) + \left[\smallunderset{JP}{\hat \Delta}{}^{uv}{}_i (\gamma, \bar{\gamma}; \tau_1) - \smallunderset{JP}{\hat \Delta}{}^{uv}{}_i (\gamma, \gamma; \tau_1)\right] \smallunderset{PP}{\Delta}^i{}_u (\gamma; \tau_1) \nonumber \\
    &\hspace{1em}+ \smallunderset{JP}{\hat \Delta}{}^{uv}{}_i (\gamma, \bar{\gamma}; \tau_1) \smallunderset{PP}{\zero \Omega}{}^i{}_u (\gamma, \bar{\gamma}; \tau_1) + 2 \smallunderset{JJ}{\hat \Delta}{}^{uv}{}_{ui} (\gamma, \bar{\gamma}; \tau_1) \smallunderset{JP}{\zero \Omega}{}^{ui}{}_u (\gamma, \bar{\gamma}; \tau_1), \\
    \smallunderset{JP}{\half \Omega}{}^{uv}{}_i (\gamma, \bar{\gamma}; \tau_1) &= 2 \smallunderset{JJ}{\Xi}^{uv}{}_{ui} (\gamma, \bar{\gamma}; \tau_1) \smallunderset{JP}{\Delta}{}^{uj}{}_i (\gamma; \tau_1) + \left[\smallunderset{JP}{\hat \Delta}{}^{uv}{}_j (\gamma, \bar{\gamma}; \tau_1) - \smallunderset{JP}{\hat \Delta}{}^{uv}{}_j (\gamma, \gamma; \tau_1)\right] \left[\delta^j{}_i + \smallunderset{PP}{\Delta}^j{}_i (\gamma; \tau_1)\right], \\
    \smallunderset{JP}{\half \Omega}{}^{vi}{}_u (\gamma, \bar{\gamma}; \tau_1) &= \smallunderset{JP}{\zero \Omega}{}^{vi}{}_v (\gamma, \bar{\gamma}; \tau_1) \smallunderset{PP}{\Delta}^v{}_u (\gamma; \tau_1) + \smallunderset{JP}{\Xi}^{vi}{}_u (\gamma, \bar{\gamma}; \tau_1) + \smallunderset{JP}{\Xi}{}^{vi}{}_j (\gamma, \bar{\gamma}; \tau_1) \smallunderset{PP}{\Delta}^i{}_u (\gamma; \tau_1) + 2 \smallunderset{JJ}{\Xi}^{vi}{}_{uj} (\gamma, \bar{\gamma}; \tau_1) \smallunderset{JP}{\Delta}^{uj}{}_u (\gamma; \tau_1) \nonumber \\
    &\hspace{1em}+ 2 \smallunderset{JJ}{\zero \Omega}{}^{vi}{}_{uv} (\gamma, \bar{\gamma}; \tau_1) \smallunderset{JP}{\Delta}^{uv}{}_u (\gamma, \bar{\gamma}; \tau_1) + \smallunderset{JJ}{\zero \Omega}{}^{vi}{}_{jk} (\gamma, \bar{\gamma}; \tau_1) \smallunderset{JP}{\Delta}^{jk}{}_u (\gamma; \tau_1) + \smallunderset{JP}{\hat \Delta}{}^{vi}{}_j (\gamma, \bar{\gamma}; \tau_1) \smallunderset{PP}{\zero \Omega}{}^i{}_u (\gamma, \bar{\gamma}; \tau_1) \nonumber \\
    &\hspace{1em}+ 2 \smallunderset{JJ}{\hat \Delta}{}^{vi}{}_{uj} (\gamma, \bar{\gamma}; \tau_1) \smallunderset{JP}{\zero \Omega}{}^{uj}{}_u (\gamma, \bar{\gamma}; \tau_1), \displaybreak[0] \\
    \smallunderset{JP}{\half \Omega}{}^{vi}{}_v (\gamma, \bar{\gamma}; \tau_1) &= \smallunderset{JP}{\zero \Omega}{}^{vi}{}_v (\gamma, \bar{\gamma}; \tau_1), \\
    \smallunderset{JP}{\half \Omega}{}^{vi}{}_j (\gamma, \bar{\gamma}; \tau_1) &= \smallunderset{JP}{\Xi}^{vi}{}_k (\gamma, \bar{\gamma}; \tau_1) \left[\delta^k{}_j + \smallunderset{PP}{\Delta}^k{}_j (\gamma; \tau_1, \tau_0)\right] + 2 \smallunderset{JJ}{\Xi}^{vi}{}_{uk} (\gamma, \bar{\gamma}; \tau_1) \smallunderset{JP}{\Delta}^{uk}{}_j (\gamma; \tau_1) \nonumber \\
    &\hspace{1em}+ 2 \smallunderset{JJ}{\zero \Omega}{}^{vi}{}_{uv} (\gamma, \bar{\gamma}; \tau_1) \smallunderset{JP}{\Delta}^{uv}{}_j (\gamma; \tau_1) + \smallunderset{JP}{\zero \Omega}{}^{vi}{}_v (\gamma, \bar{\gamma}; \tau_1) \smallunderset{PP}{\Delta}^v{}_j (\gamma; \tau_1) + \smallunderset{JJ}{\zero \Omega}{}^{vi}{}_{kl} (\gamma, \bar{\gamma}; \tau_1) \smallunderset{JP}{\Delta}^{kl}{}_j (\gamma; \tau_1, \tau_0), \\
    \smallunderset{JP}{\half \Omega}{}^{ij}{}_u (\gamma, \bar{\gamma}; \tau_1) &= \smallunderset{JP}{\Xi}^{ij}{}_u (\gamma, \bar{\gamma}; \tau_1) + \smallunderset{JP}{\Xi}^{ij}{}_k (\gamma, \bar{\gamma}; \tau_1, \tau_0) \smallunderset{PP}{\Delta}^k{}_u (\gamma; \tau_1) + 2 \smallunderset{JJ}{\Xi}^{ij}{}_{uk} (\gamma, \bar{\gamma}; \tau_1) \smallunderset{JP}{\Delta}^{uk}{}_u (\gamma; \tau_1) \nonumber \\
    &\hspace{1em}+ \smallunderset{JP}{\hat \Delta}{}^{ij}{}_k (\gamma, \bar{\gamma}; \tau_1, \tau_0) \smallunderset{PP}{\zero \Omega}{}^k{}_u (\gamma, \bar{\gamma}; \tau_1) + 2 \smallunderset{JJ}{\hat \Delta}{}^{ij}{}_{uk} (\gamma, \bar{\gamma}; \tau_1) \smallunderset{JP}{\zero \Omega}{}^{uk}{}_u (\gamma, \bar{\gamma}; \tau_1), \\
    \smallunderset{JP}{\half \Omega}{}^{ij}{}_k (\gamma, \bar{\gamma}; \tau_1, \tau_0) &= \smallunderset{JP}{\Xi}^{ij}{}_l (\gamma, \bar{\gamma}; \tau_1, \tau_0) \left[\delta^l{}_k + \smallunderset{PP}{\Delta}{}^l{}_k (\gamma; \tau_1, \tau_0)\right] + 2 \smallunderset{JJ}{\Xi}^{ij}{}_{ul} (\gamma, \bar{\gamma}; \tau_1) \smallunderset{JP}{\Delta}^{ul}{}_k (\gamma; \tau_1), \displaybreak[0] \\
    \smallunderset{JJ}{\half \Omega}{}^{uv}{}_{ui} (\gamma, \bar{\gamma}; \tau_1) &= \smallunderset{JJ}{\Xi}^{uv}{}_{uj} (\gamma, \bar{\gamma}; \tau_1) \left[\delta^j{}_i + 2 \smallunderset{JJ}{\Delta}^{uj}{}_{ui} (\gamma; \tau_1)\right] + \left[\smallunderset{JP}{\hat \Delta}{}^{uv}{}_j (\gamma, \bar{\gamma}; \tau_1) - \smallunderset{JP}{\hat \Delta}{}^{uv}{}_j (\gamma, \gamma; \tau_1)\right] \smallunderset{PJ}{\Delta}^j{}_{ui} (\gamma; \tau_1), \\
    \smallunderset{JJ}{\half \Omega}{}^{vi}{}_{uv} (\gamma, \bar{\gamma}; \tau_1) &= \smallunderset{JJ}{\zero \Omega}{}^{vi}{}_{uv} (\gamma, \bar{\gamma}; \tau_1), \\
    \smallunderset{JJ}{\half \Omega}{}^{vi}{}_{uj} (\gamma, \bar{\gamma}; \tau_1) &= \smallunderset{JP}{\Xi}{}^{vi}{}_k (\gamma, \bar{\gamma}; \tau_1) \smallunderset{PJ}{\Delta}^k{}_{uj} (\gamma; \tau_1) + \smallunderset{JJ}{\Xi}^{vi}{}_{uk} (\gamma, \bar{\gamma}; \tau_1) \left[\delta^k{}_j + 2 \smallunderset{JJ}{\Delta}{}^{uk}{}_{uj} (\gamma; \tau_1)\right] \nonumber \\
    &\hspace{1em}+ \smallunderset{JP}{\zero \Omega}{}^{vi}{}_v (\gamma, \bar{\gamma}; \tau_1) \smallunderset{PJ}{\Delta}^v{}_{uj} (\gamma; \tau_1) + \smallunderset{JJ}{\zero \Omega}{}^{vi}{}_{kl} (\gamma, \bar{\gamma}; \tau_1) \smallunderset{JJ}{\Delta}{}^{kl}{}_{uj} (\gamma; \tau_1) + 2 \smallunderset{JJ}{\zero \Omega}{}^{vi}{}_{uv} (\gamma, \bar{\gamma}; \tau_1) \smallunderset{JJ}{\Delta}^{uv}{}_{uj} (\gamma; \tau_1), \\
    \smallunderset{JJ}{\half \Omega}{}^{vi}{}_{jk} (\gamma, \bar{\gamma}; \tau_1) &= \smallunderset{JJ}{\zero \Omega}{}^{vi}{}_{jk} (\gamma, \bar{\gamma}; \tau_1), \\
    \smallunderset{JJ}{\half \Omega}{}^{ij}{}_{uk} (\gamma, \bar{\gamma}; \tau_1) &= \smallunderset{JP}{\Xi}^{ij}{}_l (\gamma, \bar{\gamma}; \tau_1, \tau_0) \smallunderset{PJ}{\Delta}^l{}_{uk} (\gamma; \tau_1) + 2 \smallunderset{JJ}{\Xi}^{ij}{}_{ul} (\gamma, \bar{\gamma}; \tau_1) \left[\delta^l{}_k + 2 \smallunderset{JJ}{\Delta}^{ul}{}_{uk} (\gamma; \tau_1)\right].
  \end{align}
\end{subequations}
\end{widetext}
This observable possesses 31 nonzero components, which is fewer than the 50 that are required by the existence of a five-dimensional space of Killing vector fields.

\section{Modifications Due to Acceleration} \label{app:acceleration}

Throughout the body of this paper, except for our spinning test particle observable, we have only considered the case where all curves used to define the observables are geodesic.
In that case, our results are nonperturbative in these spacetimes.
We now consider the effects of acceleration.
In essence, the only change is that now our results are necessarily perturbative, but fortunately only perturbative in the acceleration, not in the initial separation or relative velocity.
Moreover, unlike the case where the curves are geodesic, our final results depend on integrals, not just sums and products, of the transverse Jacobi propagators.
Note that this also seems to be the case with the spinning test particle observable, which was necessarily defined using an accelerated curve.

To compute our observables for accelerated curves, note that, given a curve $\gamma$ and a proper time $\tau_0$, there is a unique geodesic $\hat{\gamma}$ that intersects $\gamma$ at $\tau_0$ that has the same four-velocity at that point.
If an observable is defined with respect to two accelerated curves $\gamma$ and $\bar{\gamma}$, we show that this observable can be written in terms of $\hat{\gamma}$ and $\hat{\bar \gamma}$.

This process is most easily done for the holonomy observable.
First, we consider the case of the holonomy with respect to parallel transport with the metric-compatible connection.
Using the fact that the initial and final regions are flat, we find that

\begin{equation} \label{eqn:different_curve_holonomy}
  \begin{split}
    \Lambda^a{}_b (\gamma; \bar{\gamma}; \tau_1) = g^a{}_{\bar a} &\Lambda^{\bar a}{}_{\bar c} (\hat{\bar \gamma}, \bar{\gamma}; \tau_1) g^{\bar c}{}_c \\
    \times &\Lambda^c{}_d (\hat{\gamma}, \hat{\bar \gamma}; \tau_1) \left(\Lambda^{-1}\right){}^d{}_b (\hat{\gamma}, \gamma; \tau_1).
  \end{split}
\end{equation}
The holonomy $\Lambda^a{}_b (\hat{\gamma}, \hat{\bar \gamma}; \tau_1)$ has already been computed in the body of the paper in Sec~\ref{sec:holonomy}.
For the other two holonomies, we can show that

\begin{equation} \label{eqn:accelerated_holonomy}
  \left(\Lambda^{\pm 1}\right){}^a{}_b (\hat{\gamma}, \gamma; \tau_1) = \delta^a{}_b \pm \pb{\ddot \gamma} \Omega^a{}_b (\hat{\gamma}, \gamma; \tau_1) + O(\ddot{\boldsymbol \gamma}^2),
\end{equation}
where

\begin{widetext}
\begin{equation} \label{eqn:accelerated_holonomy_diff}
  \begin{split}
    \pb{\ddot \gamma} \Omega^a{}_b (\hat{\gamma}, \gamma; \tau_1) = \int_{\tau_0}^{\tau_1} \ud \tau_2 \bigg\{&(\tau_0 - \tau_2) \pb{\hat \gamma} H^c{}_{\hat c''} \Big[\pb{\xi} \Omega^a{}_{bc} (\hat{\gamma}; \tau_1) - \pb{\xi} \Omega^a{}_{bc} (\hat{\gamma}; \tau_2)\Big] \\
    &+ \frac{\uD}{\ud \tau_0} \left[(\tau_0 - \tau_2) \pb{\hat \gamma} H^c{}_{\hat c''}\right] \Big[\pb{\dot \xi} \Omega^a{}_{bc} (\hat{\gamma}; \tau_1) - \pb{\dot \xi} \Omega^a{}_{bc} (\hat{\gamma}; \tau_2)\Big]\bigg\} g^{\hat c''}{}_{c''} \ddot{\gamma}^{c''}.
  \end{split}
\end{equation}
\end{widetext}
When considering the holonomy of linear and angular momentum for a specific value of $\varkappa$, analogous versions of Eqs.~\eqref{eqn:different_curve_holonomy},~\eqref{eqn:accelerated_holonomy}, and~\eqref{eqn:accelerated_holonomy_diff} hold.

At this point, we note that $\pb{\xi} \Omega^a{}_{bc} (\hat{\gamma}; \tau_1)$ and $\pb{\dot \xi} \Omega^a{}_{bc} (\hat{\gamma}; \tau_1)$ (along with their analogues for the linear and angular momentum transport) are given in the body of the paper in Sec.~\ref{sec:holonomy}.
However, their versions with $\tau_2$ instead of $\tau_1$ [that is, $\pb{\xi} \Omega^a{}_{bc} (\hat{\gamma}; \tau_2)$ and $\pb{\dot \xi} \Omega^a{}_{bc} (\hat{\gamma}; \tau_2)$] are not, for any arbitrary value of $\tau_2 \in (\tau_0, \tau_1)$.
As the expressions for these quantities are lengthy, we merely note that they can be determined using the expressions in~\cite{Flanagan2019}, once adapted to plane wave spacetimes in Brinkmann coordinates.

We now consider the curve deviation observable, starting with the various pieces that go into the definition of the curve deviation in Eq.~\eqref{eqn:curve_deviation_def}.
Since it is crucial in this case, we make the dependence on the curves of the separation and the curve deviation observable explicit.
First,

\begin{equation}
  \xi^{a'} (\gamma, \bar{\gamma}) = g^{a'}{}_{\hat a'} \left[\xi^{\hat a'} (\hat{\gamma}, \hat{\bar \gamma}) + g^{\hat a'}{}_{\hat{\bar a}'} \xi^{\hat{\bar a}'} (\hat{\bar \gamma}, \bar{\gamma}) - \xi^{\hat a'} (\hat{\gamma}, \gamma)\right].
\end{equation}
Using the holonomy, we also find

\begin{equation}
  \pb{\gamma} g^{a'}{}_a = g^{a'}{}_{\hat a'} \pb{\hat \gamma} g^{\hat a'}{}_b \left(\Lambda^{-1}\right){}^b{}_a (\hat{\gamma}, \gamma; \tau_1),
\end{equation}
and moreover

\begin{equation}
  \begin{split}
    \pb{\gamma} g^{a'}{}_{a'''} = g^{a'}{}_{\hat a'} \pb{\hat \gamma} g^{\hat a'}{}_a &\left(\Lambda^{-1}\right){}^a{}_c (\hat{\gamma}, \gamma; \tau_1) \\
    \times &\Lambda^c{}_b (\hat{\gamma}, \gamma; \tau_3) \pb{\hat \gamma} g^b{}_{\hat a'''} g^{\hat a'''}{}_{a'''}.
  \end{split}
\end{equation}
Putting this all together, one finds that

\begin{widetext}
\begin{equation}
  \begin{split}
    \Delta \xi^{a'}_{\textrm{CD}} (\gamma, \bar{\gamma}) = g^{a'}{}_{\hat a'} \Bigg\{&\Delta \xi^{\hat a'}_{\textrm{CD}} (\hat{\gamma}, \hat{\bar \gamma}) + g^{\hat a'}{}_{\hat{\bar a}'} \xi^{\hat{\bar a}'} (\hat{\bar \gamma}, \bar{\gamma}) - \xi^{\hat a'} (\hat{\gamma}, \gamma) + \pb{\hat \gamma} g^{\hat a'}{}_a \left[\left(\Lambda^{-1}\right){}^a{}_b (\hat{\gamma}, \gamma; \tau_1) - \delta^a{}_b\right] \left[\xi^b + (\tau_1 - \tau_0) \dot{\xi}^b\right] \\
    &+ \pb{\hat \gamma} g^{\hat a'}{}_a \left(\Lambda^{-1}\right){}^a{}_c (\hat{\gamma}, \gamma; \tau_1) \int_{\tau_0}^{\tau_1} \ud \tau_2 \int_{\tau_0}^{\tau_2} \ud \tau_3 \Lambda^c{}_b (\hat{\gamma}, \gamma; \tau_3) \pb{\hat \gamma} g^b{}_{\hat b'''} g^{\hat b'''}{}_{b'''} \left(g^{b'''}{}_{\bar b'''} \ddot{\bar \gamma}^{\bar b'''} - \ddot{\gamma}^{b'''}\right)\Bigg\}.
  \end{split}
\end{equation}
\end{widetext}
The terms with holonomies in this expression are given by Eqs.~\eqref{eqn:accelerated_holonomy} and~\eqref{eqn:accelerated_holonomy_diff}.
The remaining terms are determined by noting that

\begin{equation}
  \xi^{\hat a'} (\hat{\gamma}, \gamma) = \int_{\tau_0}^{\tau_1} \ud \tau_2 (\tau_1 - \tau_2) \pb{\hat \gamma} H^{\hat a'}{}_{\hat a''} g^{\hat a''}{}_{a''} \ddot{\gamma}^{a''},
\end{equation}
with an analogous statement holding for $\xi^{\hat{\bar a}'} (\hat{\bar \gamma}, \bar{\gamma})$.

\bibliography{Refs}

\begin{table*}
  \centering
  \setlength{\tabcolsep}{9pt}
  \caption{\label{tab:symbols} A table of the symbols that occur frequently in the body of this paper.}
  \begin{tabular}{p{10em} p{0.47\textwidth} p{8em}} \hline
    Symbol & Explanation & Defining equation \\\hline\hline
    $u$, $v$, $x^i$ & Brinkmann coordinates of a plane wave & \eqref{eqn:metric} \\\hline
    $\underline{\mathcal A}{}_{ab}$ & Profile of the gravitational plane wave & \eqref{eqn:metric},~\eqref{eqn:A_def} \\\hline
    $\ell_a$ & Wave vector of the plane wave & \eqref{eqn:l} \\\hline
    $\underline{\epsilon}{}_{ab}$ & Volume form for surfaces of constant $u$ and $v$ & \eqref{eqn:volume} \\\hline
    $\Xi_a$ & One of a four-parameter family of Killing vectors in plane wave spacetimes & \eqref{eqn:killing} \\\hline
    $\gamma$, $\bar{\gamma}$ & Curves in plane wave spacetimes parametrized by $\tau$ & $\cdots$ \\\hline
    $\xi^a$, $\dot{\xi}^a$ & Separation and relative velocity of $\gamma$ and $\bar{\gamma}$, respectively & $\cdots$ \\\hline
    $\pb{\gamma} g^{a'}{}_a$ & Parallel propagator along $\gamma$, or unique geodesic between $x$ and $x'$ if $\gamma$ is unspecified & \eqref{eqn:parallel} \\\hline
    $\pb{\gamma} K^{a'}{}_a$, $\pb{\gamma} H^{a'}{}_a$ & Jacobi propagator along $\gamma$, or unique geodesic between $x$ and $x'$ if $\gamma$ is unspecified& \eqref{eqn:jacobi} \\\hline
    $\mat{K}^i{}_j (u', u)$, $\mat{H}^i{}_j (u', u)$ & Transverse Jacobi propagators (in the $x^i$ directions) & \eqref{eqn:2d_jacobi} \\\hline
    $\chi$ & Conserved quantity $\ell^a \dot{\gamma}_a$ & \eqref{eqn:chi} \\\hline
    $\pp{(n)} Q^{\cdots}{}_{\cdots}$ & Coefficients of an expansion of $Q^{\cdots}{}_{\cdots}$ at $n$th order in $\mat{\mathcal A}^i{}_j (u)$ & \eqref{eqn:2d_jacobi_expansion}, for example \\\hline
    $\epsilon$, $a$, $\phi$, $n$ & Parameters of an example of a wave profile & \eqref{eqn:model_waveform} \\\hline
    $\pb{+} \mat{\boldsymbol e}$, $\pb{\times} \mat{\boldsymbol e}$ & Plus and cross polarization matrices for an example of a wave profile & \eqref{eqn:plus_cross} \\\hline
    $\pb{x^k \dot{x}^l} [Q^{\cdots}{}_{\cdots}]$ & Coefficients in an expansion of $Q^{\cdots}{}_{\cdots}$ at $k$th order in $x$ and $l$th order in $\dot{x}$ & \eqref{eqn:polynomial_decomp} \\\hline
    $\Delta \xi_{\textrm{CD}}^{a'}$ & Curve deviation observable & \eqref{eqn:curve_deviation_def} \\\hline
    $\Delta K^{a'}{}_b$, $\Delta H^{a'}{}_b$, $L^{a'}{}_{bc}$, $N^{a'}{}_{bc}$, $M^{a'}{}_{bc}$ & Coefficients in the expansion of $\Delta \xi_{\textrm{CD}}^{a'}$ & \eqref{eqn:curve_dev_decomp} \\\hline
    $\varkappa$ & Set of four parameters $(\varkappa_1, \varkappa_2, \varkappa_3, \varkappa_4)$ that defines a method of linear and angular momentum transport & \eqref{eqn:transport}, \eqref{eqn:K_def} \\\hline
    $\vkappa{K}{}^a{}_{bcd}$ & Tensor constructed from $R^a{}_{bcd}$ that is parametrized by $\varkappa$ & \eqref{eqn:K_def} \\\hline
    $\vkappa{Q}{}^{\cdots}{}_{\cdots}$, $\zero{Q}{}^{\cdots}{}_{\cdots}$, $\half{Q}{}^{\cdots}{}_{\cdots}$ & Some quantity that is defined with respect to angular momentum transport using a general $\varkappa$, $\varkappa = (0, 0, 0, 0)$, and $\varkappa = (1/2, 0, 0, 0)$, respectively & $\cdots$ \\\hline
    $A$, $B$, etc. & Indices on the linear and angular momentum bundle~\cite{Flanagan2016, Flanagan2019} & $\cdots$ \\\hline
    $X^A$ & Combined linear and angular momentum vector $\binom{P^a}{J^{ab}}$ & \eqref{eqn:X_def} \\\hline
    $\smallunderset{PP}{A}^a{}_c$, $\smallunderset{PJ}{A}^a{}_{cd}$, $\smallunderset{JP}{A}^{ab}{}_c$, $\smallunderset{JJ}{A}^{ab}{}_{cd}$ & Blocks of a matrix $A^A{}_B$ that acts on $X^A$ & \eqref{eqn:mat_def} \\\hline
    $\vkappa{\Lambda}{}^A{}_B (\gamma, \bar{\gamma}; \tau_1)$ & Holonomy of linear and angular momentum transport & \eqref{eqn:Lambda_def} \\\hline
    $\vkappa{\Omega}{}^A{}_B (\gamma, \bar{\gamma}; \tau_1)$ & The above holonomy, minus the identity & \eqref{eqn:Omega_def} \\\hline
    $\Lambda^a{}_b (\gamma, \bar{\gamma}; \tau_1)$, $\Omega^a{}_b (\gamma, \bar{\gamma}; \tau_1)$ & Holonomy and holonomy minus identity for parallel transport & \eqref{eqn:parallel_Omega} \\\hline
    $\Delta \chi^a (\gamma, \bar{\gamma}; \tau_1)$ & ``Inhomogeneous generalized holonomy'' used for calculating holonomy for $\varkappa = (0, 0, 0, 0)$ & \eqref{eqn:Delta_chi} \\\hline
    $\Delta^A{}_B (\gamma; \tau_1)$, $\hat{\Delta}{}^A{}_B (\gamma, \bar{\gamma}; \tau_1)$ & Quantities which occur in the calculation of the holonomy for $\varkappa = (1/2, 0, 0, 0)$ & \eqref{eqn:Delta_defs} \\\hline
    $\Upsilon^{a'}{}_b$, $\Psi^{a'}{}_{bc}$ & Coefficients that occur in a perturbative expansion of the spinning test particle observables & \eqref{eqn:spin_obs_decomp} \\\hline
  \hline \end{tabular}
\end{table*}

\end{document}